\documentclass[aps,prb,twocolumn,showpacs,preprintnumbers,amsmath,amssymb,superscriptaddress]{revtex4-1}

\usepackage{graphicx}
\usepackage{dcolumn}
\usepackage{amsmath,mathrsfs,amssymb,amsthm}
\usepackage{hhline}
\usepackage{wasysym,enumerate,color}
\usepackage{epstopdf}
\usepackage{verbatim}
\usepackage{hyperref}
\usepackage{color}
\usepackage{ulem} % needed for \scrap-command
\renewcommand{\emph}[1]{\textit{#1}} % needed, if the package  "ulem" is used

\definecolor{darkgreen}{rgb}{0,0.5,0}
\definecolor{purple}{rgb}{0.35,0,0.35}
\definecolor{orange}{rgb}{1,0.5,0}
\definecolor{darkred}{rgb}{.7,0,0}
\definecolor{darkblue}{rgb}{0,0,.3}
\definecolor{grey}{rgb}{.6,.6,.6}
\definecolor{dimgreen}{rgb}{0.2,0.6,0.1}
\hypersetup{colorlinks,linkcolor=blue,urlcolor=blue,citecolor=blue}

\newcommand{\be}{\begin{equation}}
\newcommand{\ee}{\end{equation}}
\newcommand{\bea}{\begin{eqnarray}}
\newcommand{\eea}{\end{eqnarray}}

\newcommand{\e}{\varepsilon}

\newcommand{\s}{{\sigma}}
\newcommand{\cG}{{\cal G}}

\newcommand{\cT}{{\cal T}}
\newcommand{\cA}{{\cal A}}

\newcommand{\bM}{{\mathbf M}}
\newcommand{\br}{{\mathbf r}}
\newcommand{\bk}{{\mathbf k}}
\newcommand{\bs}{{\boldsymbol \sigma}}
\newcommand{\rM}{{\rm M}}

\newcommand{\Fk}{F({\mathbf{k}})}
\newcommand{\Gk}{G({\mathbf{k}})}

\begin{document}

\title{Probing the Rashba effect via the induced magnetization around a Kondo impurity}

\author{R. Chirla}
\email{chirlarazvan@yahoo.com}
\affiliation{Department of Physics, University of Oradea, 410087, Oradea, Romania}

\author{C. P. Moca}
\affiliation{Department of Physics, University of Oradea, 410087, Oradea, Romania}
\affiliation{BME-MTA Exotic Quantum Phase Group, Institute of Physics, Budapest University of Technology and Economics,
H-1521 Budapest, Hungary}

\author{I. Weymann}
\affiliation{Faculty of Physics, Adam Mickiewicz University, 61-614 Pozna\'n, Poland}

\date{\today}

%%%%%%%%%%%%%%%%%%%%%%%%%%%%%%%%%%%%%%%%%%%%%%%%%%%%%%%%%%%%%%%%
%%%%%%%%%%%%%%%%%%%%%%%%%%%%%%%%%%%%%%%%%%%%%%%%%%%%%%%%%%%%%%%%

\begin{abstract}
When a  single magnetic adatom is deposited on a surface of a metal,
it affects the charge and spin texture of the electron gas surrounding it.
The screening of the local moment by conduction electrons gives rise
to the Kondo effect.
Here we investigate the effect of the Rashba spin orbit
coupling on the local magnetization density of states (LMDOS)
around a Cobalt impurity on a Au(111) surface in a magnetic field.
We show that the in-plane component of the LMDOS  
is exclusively associated with the Rashba spin orbit interaction.
This observation can be experimentally exploited to confirm the presence of
the Rashba effect on surfaces, such as Au(111),
by performing spin-polarized STM measurements around the Kondo impurity.
\end{abstract}

% 72.15.Qm    Scattering mechanisms and Kondo effect
% 72.25.-b    Spin polarized transport
%75.70.Tj		Spin-orbit effects
\maketitle

%%%%%%%%%%%%%%%%%%%%%%%%%%%%%%%%%%%%%%%%%%%%%%%%%%%%%%%%%%%%%%%%
%%%%%%%%%%%%%%%%%%%%%%%%%%%%%%%%%%%%%%%%%%%%%%%%%%%%%%%%%%%%%%%%

\section{Introduction}\label{sec:Intro}

%%%%%%%%%%%%%%%%%%%%%%%%%%%%%%%%%%%%%%%%%%%%%%%%%%%%%%%%%%%%%%%%
%%%%%%%%%%%%%%%%%%%%%%%%%%%%%%%%%%%%%%%%%%%%%%%%%%%%%%%%%%%%%%%%

In two-dimensional structures, the coupling between the spin and angular momentum
can lead to a variety of interesting phenomena.~\cite{winkler}
Interestingly enough, the Rashba spin-orbit (SO) interaction\cite{rashba} has
become one of the intriguing and desired ingredients
of modern nanoelectronics and spintronics.~\cite{Awschalom.09}
Studying the Rashba-induced effects
for atoms placed on surfaces is especially interesting, because
on one hand, it opens up a way to probe the strength of the SO coupling, and
on the other hand, it may allow one to explore the interplay between the Kondo screening and the SO interaction.
In this regard, the Au(111) surface is very promising,
since it exhibits a measurable energy splitting of surface band states,
which was first experimentally observed 
by LaShell {\it et al.}~[\onlinecite{LaShell.96}].
However, because the Friedel oscillations induced by an adatom
remain practically unaffected by the Rashba SO interaction~\cite{Petersen.00},
the presence of the Rashba effect could not be verified
by typical scanning tunneling microscopy (STM) measurements,
focused mainly on the charge transport spectroscopy.
Consequently, not much attention has been paid to it so far.
It has been noticed only very recently that placing a  magnetic adatom
on the surface can help probe the Rashba spin-orbit effect.
By treating the impurity at the classical level and in the absence of the Kondo effect,
Lounis {\it et al.}~[\onlinecite{Lounis.12}] 
have shown that the induced spin polarization
of the electron gas surrounding the magnetic adatom exhibits a spin texture,
which is a superposition of two skyrmionic waves with opposite chirality.
This has been attributed to the presence of the
Rashba SO interaction.

\begin{figure}[t]
\begin{center}
\includegraphics[width=0.47\columnwidth]{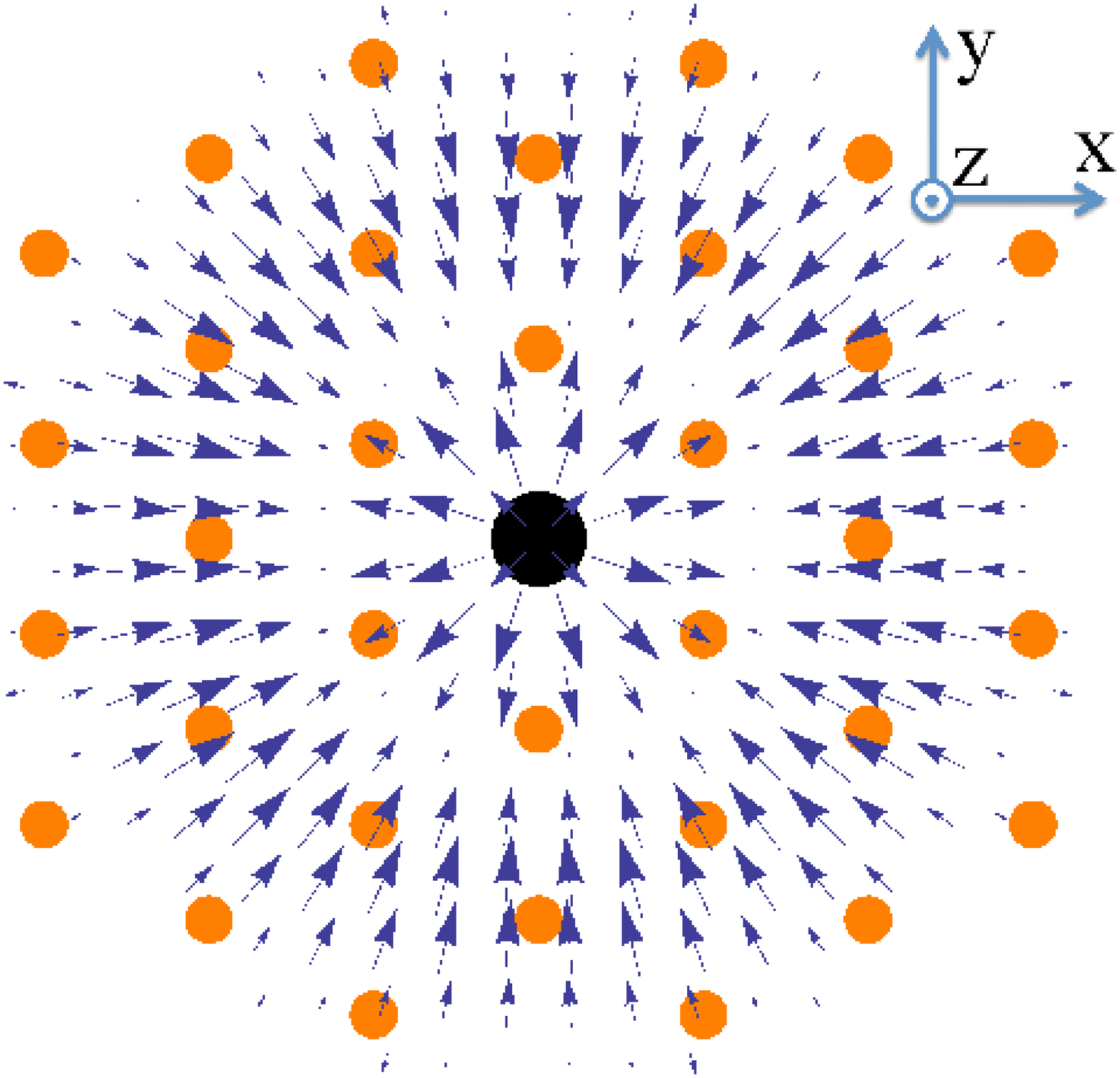}
\includegraphics[width=0.51\columnwidth]{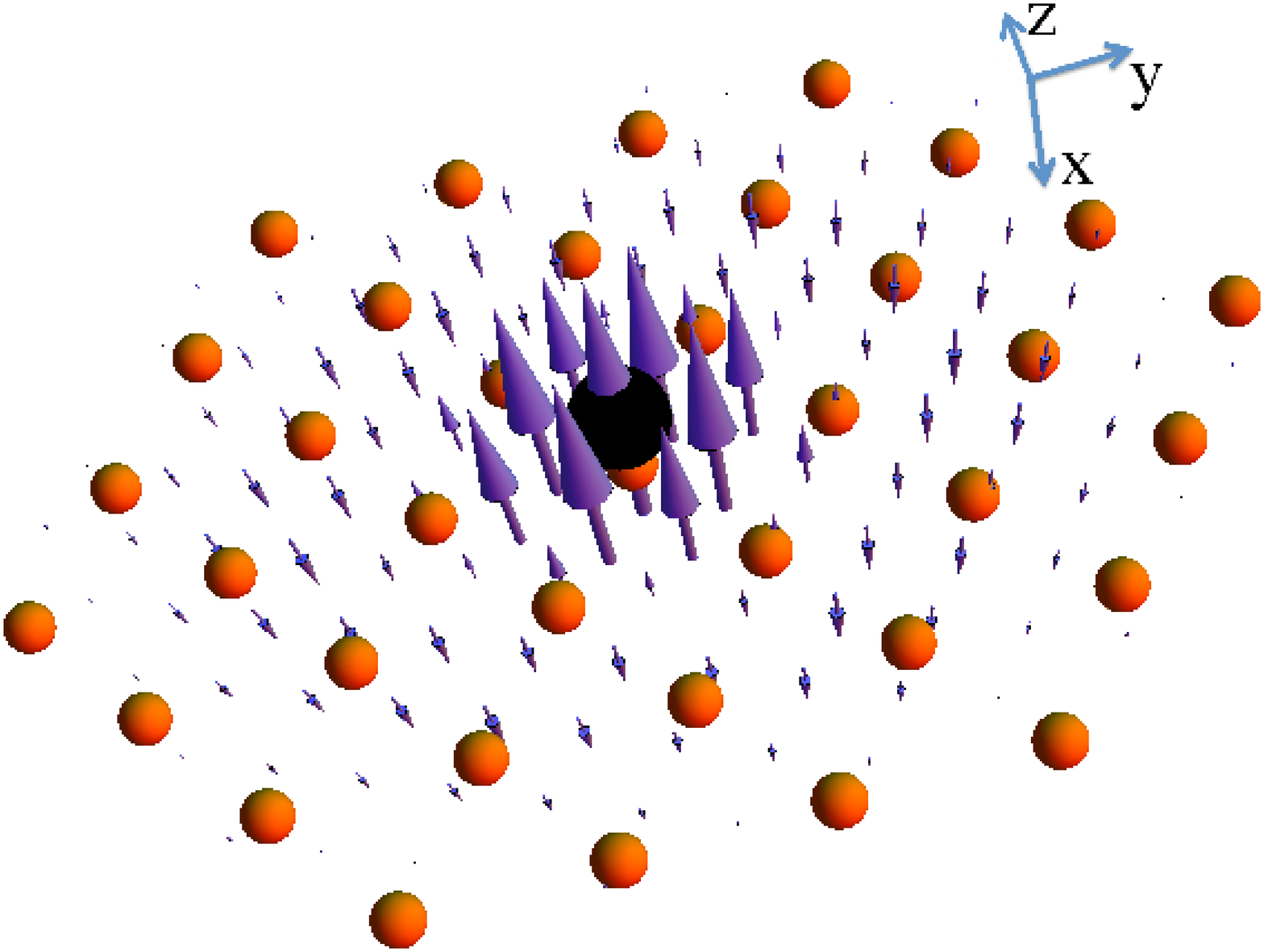}
\caption{
  (Color online) Spatial distribution of the radial component $\bM_r$ (left panel)
  and of the total LMDOS $\bM$ (right panel)
  at energy $E=-2$ meV below the Fermi surface, and for a
  magnetic field $B=3$ T applied along the $z$th direction.
  The smaller (orange) spheres indicate the position of the Au atoms on the Au(111)
  surface, and the large (black) sphere represents the magnetic adatom.
  The arrows indicate the amplitude and direction of the LMDOS.
 }
\label{fig:magnetization_2d}
\end{center}
\end{figure}

In the present work we pursue this problem further and study 
the local texture of the spin-resolved density of states around the magnetic adatom
in the case of nontrivial many-body interactions, such as the ones leading to the Kondo 
effect,~\cite{Kondo.64} and in the presence of an external magnetic field  applied along 
the $z$th direction.
The Kondo effect, which occurs when a local impurity
spin in a metallic host is screened by the conduction electrons, 
is undoubtedly one of the fundamental effects in condensed matter physics.~\cite{hewson}
Here we investigate how the induced local magnetization density of states (LMDOS)
around a magnetic adatom in the Kondo
regime, in the presence of an external magnetic field, is affected by the Rashba SO 
interaction.

In the many body formalism, the LMDOS, $\bM(\br, E)$, can be expressed
in terms of the retarded, single particle Green's function,
$\cG(\br, \br', E)$, as:
\begin{equation} \label{Eq:LMDOS}
  \bM(\br, E) = -\frac{1}{\pi}\,\Im m\, {\rm Tr}_\s \{\bs \; {\cG}(\br, \br, E)\}\,,
\end{equation}
where $\bs$ are the Pauli matrices,
$E$ is the energy measured with respect to the Fermi energy $E_F$,
and $\br$ is the in-plane distance from the impurity.
The real-space Green's function
$\cG(\br, \br, E)$ shall be computed 
in terms of the many body $\cT$-matrix for the conduction electrons,
which describes the scattering of the surface electrons off the impurity.

To make quantitative estimates, we focus on a Co atom
on a Au(111) surface,~\cite{Madhavan.98, Madhavan.01}
for which a Kondo temperature $T_K$ of about $75$ K
was extracted from STM spectroscopy measurements.~\cite{Madhavan.01}
The Au(111) surface is modeled within the tight binding approximation (TBA),
which, in spite of its simplicity, is able to properly describe the 
dispersion of the Au surface states.~\cite{Liu.08}
To capture the Kondo physics correctly, the Co impurity is described in 
terms of the Anderson model.~\cite{Anderson.61}
The many body $\cT$-matrix is related to the Green's function describing the  
local orbitals of the Co ion, (see Sec.~\ref{sec:T-matrix})
which can be then computed with the aid of the numerical renormalization
group (NRG) method,~\cite{Wilson.75}  known as the most versatile and accurate
in treating quantum impurity problems.
Moreover, to make realistic predictions,
in NRG calculations we take into account the full energy dependence
of the density of states (DOS) of the Au(111) surface.
While in general the magnetic impurity itself can have a complicated 
orbital structure, and channels with different symmetries 
may couple to the surface, within the NRG approach
the coupling is assumed to have an $s$-wave symmetry.

One of the main results of this paper--a 
non-vanishing in-plane magnetization ${\bM_\br}$--
is sketched in Fig.~\ref{fig:magnetization_2d}
together with the total magnetization $\bM $
in a magnetic field, $B=$ 3 T.
Notice that at $B=0$, the 
local polarization vanishes, $\bM=0$.
Although the total LMDOS $\bM = (\rM_x, \rM_y, \rM_z)$ 
is an important quantity, we have found that
the in-plane component $\bM_\br$ is much more interesting, as it is
strongly affected by the presence of the SO interaction.
On the other hand, the out of plane part, $\rM_z$, 
depends weakly on the Rashba interaction and displays a
spatial behavior that is somewhat similar to the one observed in
the local density of states (LDOS),~\cite{Madhavan.98, Knorr.02, Schneider.05}
More than that, $\bM _\br$ is a pure Rashba effect,
as it vanishes if the SO interaction is turned off.

Experimentally, it is possible to measure
the radial component of the energy-dependent LMDOS
with the state-of-the-art spin-polarized STM techniques,~\cite{Wiesendanger.09} 
which can thus provide an
important information on the presence
and strength of the Rashba SO interaction.
In this paper, using realistic parameters,
we study the behavior of the energy
and position-dependent LMDOS in
the region around the magnetic impurity.

The paper is organized as follows: In section \ref{sec:TF} 
we introduce our model Hamiltonian that describes the Au surface.
The description of the magnetic impurity problem is also presented
in the same section.
In section \ref{sec:NR} we present and discuss our numerical results.
We close with conclusions in \ref{sec:Conclusions}, where our main findings 
are reiterated.

%%%%%%%%%%%%%%%%%%%%%%%%%%%%%%%%%%%%%%%%%%%%%%%%%%%%%%%%%%%%%%%%
%%%%%%%%%%%%%%%%%%%%%%%%%%%%%%%%%%%%%%%%%%%%%%%%%%%%%%%%%%%%%%%%

%\section{Hamiltonian and Green's function}
\section{Theoretical framework}\label{sec:TF}

%%%%%%%%%%%%%%%%%%%%%%%%%%%%%%%%%%%%%%%%%%%%%%%%%%%%%%%%%%%%%%%%
%%%%%%%%%%%%%%%%%%%%%%%%%%%%%%%%%%%%%%%%%%%%%%%%%%%%%%%%%%%%%%%%

%\subsection{Model Hamiltonian for Au(111) surface}
\subsection{Modeling the Au(111) surface}\label{sec:surface}

%%%%%%%%%%%%%%%%%%%%%%%%%%%%%%%%%%%%%%%%%%%%%%%%%%%%%%%%%%%%%%%%
%%%%%%%%%%%%%%%%%%%%%%%%%%%%%%%%%%%%%%%%%%%%%%%%%%%%%%%%%%%%%%%%

Let us introduce the details of the lattice under investigation, the
tight-binding Hamiltonian describing it, and the corresponding band structure.
The Au(111) surface presents a hexagonal structure, with one atom per unit cell. The
basis vectors of the direct lattice are ${\bf t}_1 = (\sqrt{3}/2, 1/2)\,a$
and ${\bf t}_2 = (-\sqrt{3}/2, 1/2)\,a$, with $a$ the lattice constant ($a=5.75$ \AA).
Here we are particularly interested in the changes induced locally by a  magnetic impurity
in the LMDOS. We shall not address the so-called herringbone 
reconstruction~\cite{Chen.98}, which 
may be relevant when analyzing photoemission spectra. 
Also, the external magnetic field is assumed to produce no kinetic effects on 
the surface states, as its effect is marginal.
In spite of its simplicity, this tight binding description is rather robust,
and can be checked against more sophisticated ab-initio
band structure calculations,~\cite{Takeuri.91}
or compared to experimentally measured binding energies.~\cite{LaShell.96}

%%%%%%%%%%%%%%%%%%%%%%%%%%%%%%%%%%%%%%%%%%%%%%%%%%%%%%%%%%%%%%%%
%%%%%%%%%%%%%%%%%%%%%%%%%%%%%%%%%%%%%%%%%%%%%%%%%%%%%%%%%%%%%%%%

\subsubsection{Hamiltonian for the Au(111) surface}\label{sec:Au}

%%%%%%%%%%%%%%%%%%%%%%%%%%%%%%%%%%%%%%%%%%%%%%%%%%%%%%%%%%%%%%%%
%%%%%%%%%%%%%%%%%%%%%%%%%%%%%%%%%%%%%%%%%%%%%%%%%%%%%%%%%%%%%%%%

We model the Au(111) surface in terms of a tight binding Hamiltonian,~\cite{Liu.08}
taking into account the
hopping between the nearest-neighbor $p_z$-orbitals subject to the Rashba SO interaction
 \begin{eqnarray} \label{Eq:H0}
H_0& = &\sum_{<\br , \br'>}\sum_{\sigma }
\left( t_{\mathbf{r}, \mathbf{r'}} +E_p
  \;\delta_{\mathbf{r}, \mathbf{r'}}\right)
 \Psi^{\dagger}_{ \mathbf{r}, \sigma}
\Psi _{ \mathbf{r'}, \sigma}
\label{eq:h0} \\
& +& i~\alpha \sum_{<\br, \br'>}\sum_{\s \s'}
  \left [ \bs \times \frac{\br- \br'}{|\br- \br'|}
\right ]^z_{\s \s'}
\Psi^{\dagger}_{ \br, \s}
\Psi _{ \mathbf{r'}, \s'} \nonumber\,.
\end{eqnarray}
The first term describes the hopping and the on-site
energies, while the second one is due to the 
Rashba spin-orbit coupling. Here, $\Psi^\dagger_{\br, \s}$ 
creates an electron in the Au ${p_z}$-orbital
at position $\br$ with spin $\sigma$,
$t_{\br, \br'}$ are the nearest-neighbor hoppings
between these orbitals,
and $E_p$ denotes their on-site energies. In the second term, 
$\alpha$ is the strength of Rashba interaction.

By fitting the tight binding dispersion, $\e_{\tau}(\bk)$, to the 
experimentally measured binding energy of Ref.~[\onlinecite{LaShell.96}] along the
$\bar\Gamma\bar M $ direction, one can extract the band parameters~\cite{Liu.08}:
$t=-0.30$~eV, $E_P = 1.41$~eV and  $\alpha = 0.02 $ eV.
We note that the effect of the external magnetic field on the 
surface electrons is rather minimal.
A simple analysis of the energies involved, shows that 
the Zeeman splitting for a magnetic field of about $3$ T
(the g-factor was taken to be $g=2.5$) is $\Delta E_Z = 0.43$~meV,
i.e. five times smaller than the Rashba energy:
$E_R = -3 \alpha^2/(2 t) = 2.08$~meV,
and tiny as compared to the band parameters $t$ and $E_p$.
Consequently, the effect of the magnetic field on the conduction electrons is neglected.

% \rc{ (These last two sentences are questionable: a different g-factor = 2 should be used for conduction electrons (different from the g for Co), then $\Delta E_Z= 0.34 $~meV. Then what about the B-L interaction ? Taking m* form Liu, the Landau level splitting $\hbar eB/m^*$ is about 1.3 meV).}

%of the order of $E_Z = g \mu_B B_z \simeq 0.434$ meV for a field of
%about $3$ T and g-factor equal to 2.5.~\cite{g_factor}
%On the other hand, the Rashba energy $E_R = 2.08$ meV for Au surface,
%which is at least one order of magnitude larger.

The Hamiltonian (\ref{Eq:H0}) can be diagonalized in Fourier space by expanding
the field operators $\Psi _{ \br, \sigma}$ as
\begin{equation}
\Psi _{\br,\sigma}  =\frac{1}{\sqrt{\Omega}}
                     \sum\limits_{\bk, \tau}
                                      e^{i\mathbf{k\; r}}
                                     \; e_{\tau,\s}\left( \mathbf k \right )
                                      \;c_{\mathbf{k}, \tau}\;, \label{eq:c}
\end{equation}
where $\Omega$ is the number of unit cells, and $ c_{\mathbf{k}, \tau}$ annihilates
an electron with momentum $\bk$ in the chiral band $\tau=\pm 1$.
Then, the dispersion is:
$\varepsilon_{\tau}(\bk) = E_p+ \Gk +\tau \,  |\Fk | $, with 
\begin{multline}
\Gk = 2t \left [ 2\cos \left (\frac{\sqrt{3}}{2} k_x a \right )
\cos \left ( \frac{1}{2} k_x a \right ) +\cos (k_y a) \right ] ,\\
\Fk = - \alpha  \left[(1+\sqrt{3} i)\sin \left ( \frac{\sqrt{3}}{2}k_x a +\frac{1}{2} k_y a \right )+\right .
\\
\left .+(1-\sqrt{3} i)\sin \left (\frac{-\sqrt{3}}{2}k_x a+\frac{1}{2} k_y a \right ) +2\sin k_y a  \right ]\, .\\
\end{multline}
Here, the form factor $\Gk$ is coming from the nearest-neighbor hopping
and $\Fk$ is due to the Rashba SO interaction.
The chiral band energies $\varepsilon_{\tau}(\bk)$, and the wave function amplitudes
 $e_{\tau, \s } (\bk)$ are determined by solving the eigenvalue equation
\begin{equation}
\sum_{\s'}H_{\s\s'}\left(
\mathbf{k} \right)\; e_{\tau, \s'}(\bk)=
\varepsilon_{\tau}(\bk)\; e_{\tau, \s} (\bk)\;,
\label{eq:Hamiltonian_matrix}
\end{equation}
with $H_{\s\s'}( \mathbf{k})$ the components of the Hamiltonian
\eqref{eq:h0} in Fourier space and spin basis.
Its matrix form is
\begin{eqnarray}
H(\bk)=
\left(
\begin{array}{cc}
E_p+\Gk & \Fk \\
\Fk^* & E_p+\Gk
\end{array}
\right).
\end{eqnarray}
The corresponding eigenvectors can be evaluated analytically,
\begin{eqnarray}
e_{+}(\bk) &=&
\frac{1}{\sqrt{2}}
\left(
1,
\frac{\Fk^*}{|\Fk|}
\right)^{\!\!T}, \nonumber\\
e_{-}(\bk) &=&
\frac{1}{\sqrt{2}}
\left(
1,
-\frac{\Fk^*}{|\Fk |}
\right)^{\!\!T} .
\label{eq:eigenvectors}
\end{eqnarray}
Then, in terms of the creation and annihilation operators for electrons
in the chiral basis, $H_0$ becomes
\begin{equation} \label{eq:H_0}
H_0 = \sum_{\bk,\tau}\, \varepsilon_\tau(\bk)
\;c^{\dagger}_{\bk,\tau}
\,c_{\bk,\tau} \,,
\end{equation}
with the operators $c_{\bk,\tau}$ satisfying the canonical anti-commutation relations,
$ \{ c_{\bk,\tau}, c^{\dagger}_{\bk',\tau'} \}
= \delta(\bk-\bk')\delta_{\tau,\tau'} $.
In this way, the surface can be described in terms of free states,
but with some chiral band structure.

%%%%%%%%%%%%%%%%%%%%%%%%%%%%%%%%%%%%%%%%%%%%%%%%%%%%%%%%%%%%%%%%
%%%%%%%%%%%%%%%%%%%%%%%%%%%%%%%%%%%%%%%%%%%%%%%%%%%%%%%%%%%%%%%%

\subsubsection{Non-interacting Green's function}

%%%%%%%%%%%%%%%%%%%%%%%%%%%%%%%%%%%%%%%%%%%%%%%%%%%%%%%%%%%%%%%%
%%%%%%%%%%%%%%%%%%%%%%%%%%%%%%%%%%%%%%%%%%%%%%%%%%%%%%%%%%%%%%%%

In the non-interacting limit, the retarded Green's function in the real space
is defined in terms of the field operators for the conduction electrons as,
\begin{equation} \label{Eq:G0}
  \cG^{(0)}_{\sigma\sigma'}(\br_i, t\,; \br_j, t') =
  -i \theta(t-t')\langle \{
  \Psi _{\br_i,\sigma} (t),
  \Psi^{\dagger} _{\br_j,\sigma'} (t')\}
  \rangle _{0} ,
\end{equation}
where $\theta(t)$ is the Heaviside function. 
Because of the spin-orbit interaction,
$\cG^{(0)}_{\sigma\sigma'}(\br_i, t\,; \br_j, t')$ is diagonal in the chiral,
but not in the spin space.
In this context, the Fourier transformed, 
non-interacting retarded Green's function for the chiral band
$\tau$ is: 
$\cG^{(0)}_{\tau}(\bk,\omega) = (\omega-\varepsilon_{\tau}(\bk) +i0^+)^{-1}$.
Then, transforming back to the spin space one gets
\begin{equation}
  \cG^{(0)}_{\s\s'} (\br_i,\br_j, \omega) =
  \sum_{\bk,\tau}\frac{e_{\tau,\sigma}(\bk)e^{*}_{\tau,\sigma'}(\bk)e^{i\bk(\br_i-\br_j)}}
  {\omega-\varepsilon_{\tau}(\bk)+i0^{+}}.
  \label{eq:green_non_interaction}
\end{equation}
This expression allows us to evaluate the density of states
(DOS) for the conduction electrons as:
\begin{eqnarray}
\varrho^{(0)}(\omega) &  = &  -\frac{1}{\pi}{\rm Tr}_{\sigma} \; {
\Im m\, \cG^{(0)}_{\s\s} (\br_i, \br_i, \omega)}.\label{eq:dos}
\end{eqnarray}

\begin{figure}[t]
\includegraphics[width=1.0\columnwidth]{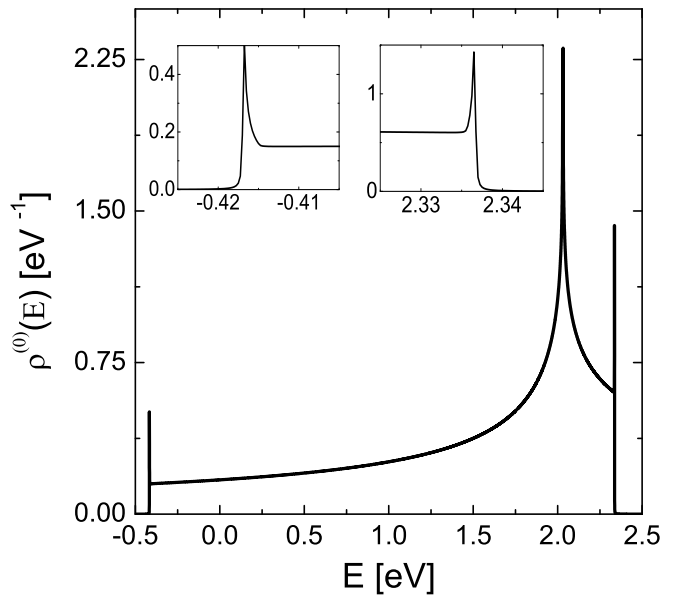}
\caption{The normalized DOS for the Au(111) surface
obtained within the TBA. The two insets show
the singularities at the band tails due to the Rashba SO interaction.
}
\label{fig:non_interacting_dos}
\end{figure}

In Fig.~\ref{fig:non_interacting_dos} we show the DOS
computed by using Eq.~\eqref{eq:dos} for the Au(111) surface.
It displays a large van Hove feature at $\omega \approx 2.03$~eV
and two sharp singularities at the band tails,
see the insets in Fig.~\ref{fig:non_interacting_dos}.
The latter features are induced by the presence of the Rashba SO interaction.
This DOS will be the input for the NRG calculations
when solving the quantum impurity problem.

%%%%%%%%%%%%%%%%%%%%%%%%%%%%%%%%%%%%%%%%%%%%%%%%%%%%%%%%%%%%%%%%
%%%%%%%%%%%%%%%%%%%%%%%%%%%%%%%%%%%%%%%%%%%%%%%%%%%%%%%%%%%%%%%%

\subsection{Modeling the quantum impurity}

%%%%%%%%%%%%%%%%%%%%%%%%%%%%%%%%%%%%%%%%%%%%%%%%%%%%%%%%%%%%%%%%
%%%%%%%%%%%%%%%%%%%%%%%%%%%%%%%%%%%%%%%%%%%%%%%%%%%%%%%%%%%%%%%%

\subsubsection{Impurity Hamiltonian}

%%%%%%%%%%%%%%%%%%%%%%%%%%%%%%%%%%%%%%%%%%%%%%%%%%%%%%%%%%%%%%%%
%%%%%%%%%%%%%%%%%%%%%%%%%%%%%%%%%%%%%%%%%%%%%%%%%%%%%%%%%%%%%%%%

To carry out the quantitative analysis of our magnetic impurity problem, 
we first need to establish how the magnetic ion couples to the
chiral bands. We consider here the {\it top configuration}, in which one Co 
atom is located on top of an Au atom, to which it hybridizes.
%
%Irek: I removed TC. I have a feeling that we introduce to many abbreviations,
% and because in text we refer to TC only two times, I think we can write explicitly "top configuration"
% this will more clear to the reader
%
The hybridization with all the other neighboring Au atoms is neglected. We have also considered other geometrical configurations
(results are not presented here),  where for example the Co atom is placed in plane,
in the middle of a hexagon,  or substitutes an Au atom, hybridizing with the nearest 
neighbors. Despite its simplicity, the considered top configuration
captures entirely the essential physics.
Correspondingly, the hybridization Hamiltonian is written as
\be
H_{V} =V \sum_{\s} \left (\;\Psi ^{\dagger }_{\br_{\rm imp}, \sigma} d_{\sigma} + h.c.\;\right ).
\label{eq:H_V1}
\ee
Here $\br_{\rm imp}$ labels the Au site below the Co impurity,
$d_{\s}$ annihilates an electron with spin $\s$ at the Co orbital, and
$V$ denotes the hopping between the two orbitals.
Within  the NRG approach, it is convenient to model the 
magnetic impurity by a single local orbital which carries only a spin label.
Transformed to the chiral basis, Eq.~\eqref{eq:H_V1} becomes
\begin{equation}
H_{V} = \sum_{\tau = \pm}\sum_{\bk,\s}\left (V_{\tau,\s}(\bk) \;c^{\dagger }_{\bk,\tau}  d
_{\sigma} + h.c.\;\right).
\label{eq:H_V}
\end{equation}
This expression is quite general, and the particular location
of the impurity atom is reflected in the $\bk$-dependence of the
hybridization factor, $V_{\tau, \s}(\bk)$. 
For the top configuration, one has
$V_{\tau, \s}(\bk)= V\, e_{\tau, \s}^* (\bk)$.
Using Eq.~\eqref{eq:eigenvectors} for the eigenvectors, one then finds:
$V_{\tau,\uparrow}(\bk) =  V/\sqrt{2} $ and
$V_{\tau, \downarrow}(\bk) =  \tau\,V\, \Fk/(\sqrt{2}|\Fk|)$.
With the hybridization Hamiltonian \eqref{eq:H_V} at hand, the total Hamiltonian that
describes the Co ion itself and the hybridization to the surface is:
\begin{equation}
H = H_{\rm imp} +H_V\; ,
\end{equation}
with
\begin{equation}
H_{\rm imp} = \sum_{\s}\varepsilon_{d\s}d^{\dagger}_{\sigma}d_{\sigma}
+ U n_{\uparrow}n_{\downarrow}\;.
\label{eq:H_imp}
\end{equation}

This Hamiltonian is similar to the single-impurity Anderson Hamiltonian,~\cite{Anderson.61}
but with a somewhat modified hybridization.
The first term in Eq.~(\ref{eq:H_imp}) describes the on-site energy 
$\e_{d\s} = \e_d \pm \frac{1}{2} g\mu_B B$ of the localized orbital,
where we included a Zeeman splitting term due to the external magnetic field $B$
applied along the $z$th direction.
We assume that the g-factor for the Co atom on the Au(111) surface is around $g=2.5$.
In the second term, $U$ represents the Coulomb repulsion
felt when two electrons with opposite spins occupy the orbital,
with $n_{\sigma} = d^{\dagger}_{\sigma}d_{\sigma}$ denoting the occupation number.
We take $U$ and $\e_d$ from ab-initio calculations~\cite{Ujsaghy.00}:
$\e_d = -0.84$ eV and $U= 2.85$ eV.
The hybridization amplitude $V$ is fixed by the Kondo temperature
itself. Here, we define the Kondo temperature $T_K$ as
the half width at half maximum (HWHM) of the
spectral function for the local orbital operator $d_\s$ in the absence of 
an external magnetic field.
Then, to get $T_K=75$ K, we take $V = 0.25$ eV.

In the presence of a $SU(2)$ spin symmetry,
the electrons in the spin=$\{\uparrow, \downarrow \}$ channels are 
scattered in the same way by the magnetic impurity.
In order to observe any spin-resolved signal,
it is necessary to break this symmetry
by applying an external magnetic field along the $z$th direction.
Then, the scattering becomes spin dependent,
as the Kondo resonance is spin-split.~\cite{Costi.2000,Kretinin.2011}
%
% Irek: I think the scattering is spin-dependent also at the Fermi level
%
One drawback of such a set-up is due to the  
large Kondo temperature~\cite{Wei.89}:
a relatively large magnetic field is necessary in order to produce a detectable 
splitting of the Kondo resonance. Here we have considered $B=3 \:$T.

%%%%%%%%%%%%%%%%%%%%%%%%%%%%%%%%%%%%%%%%%%%%%%%%%%%%%%%%%%%%%%%%
%%%%%%%%%%%%%%%%%%%%%%%%%%%%%%%%%%%%%%%%%%%%%%%%%%%%%%%%%%%%%%%%
\subsubsection{Calculation of the $\cT$-matrix}\label{sec:T-matrix}
%%%%%%%%%%%%%%%%%%%%%%%%%%%%%%%%%%%%%%%%%%%%%%%%%%%%%%%%%%%%%%%%
%%%%%%%%%%%%%%%%%%%%%%%%%%%%%%%%%%%%%%%%%%%%%%%%%%%%%%%%%%%%%%%%

To solve the quantum impurity problem, we employ Wilson's NRG method.~\cite{Wilson.75}
NRG is a powerful tool for accurate calculations of equilibrium properties of arbitrarily complex quantum 
impurities coupled to electron reservoirs.~\cite{Bulla.2008}
The method consists in the logarithmic discretization of the continuum of conduction 
states, followed by a mapping to a one dimensional chain Hamiltonian (Wilson chain)
with exponentially decaying hoppings.
The mapping starts with expanding the operators $c_{\bk, \tau}$
in terms of the eigenfunctions of the angular momentum~\cite{Krishnamurthy.80},
\begin{equation}
c_{\bk, \tau} = \frac{1}{\sqrt{ |k| }}\;
\sum_{m=-\infty}^{\infty}
\frac{1}{\sqrt{2\pi}} e^{i m\phi}c_{k,\tau}^{m}\,
\end{equation}
and then constructing an effective impurity model
by integrating out the electronic angular momentum modes.
The broadening felt by the impurity is given by the imaginary part of the hybridization 
function
\begin{eqnarray}
\Delta_{\s}(\omega)& =&\;\sum_{\tau}\sum_\bk\,
\frac{|V_{\tau,\s}(\bk)|^2}{\omega-\e_{\tau}(\bk)+i 0^{+}} .
\end{eqnarray}
To a first approximation, the Rashba spin orbit coupling is weak and 
does not affect the impurity spectral function:  
$\Delta_{\uparrow}(\omega) = \Delta_{\downarrow}(\omega)$.

\begin{figure}[h]
\includegraphics[width=0.9\columnwidth]{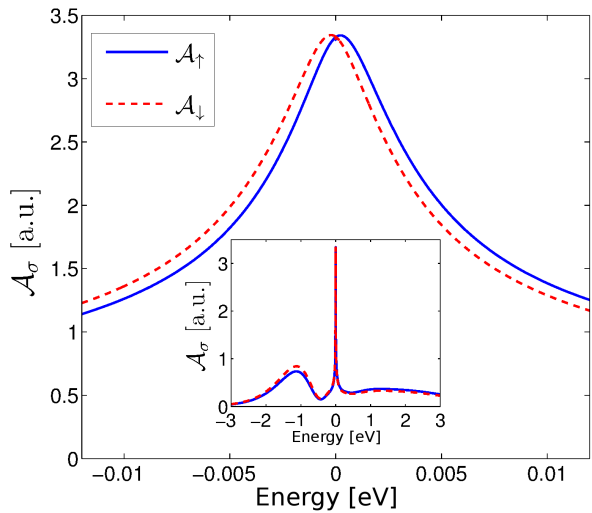}
\caption{(Color online) The energy dependence of the spin-dependent
spectral function $\cA_{\s}(\omega)$ of the local orbital
around the Fermi level, calculated within the NRG approach.
The splitting of the Kondo resonance
is due to the external magnetic field, $B = 3$ T.
The inset shows the full energy dependence of $\cA_{\s}(\omega)$.
The parameters for the Anderson model are: $V=0.25$ eV,
$U = 2.85$ eV and $\varepsilon_{d} = -0.84$ eV.}
\label{fig:spectral_function}
\end{figure}

In order to determine the LMDOS, Eq.~(\ref{Eq:LMDOS}),
one needs to calculate the full Green's function $\cG_{\s\s'}(\br; \br; \omega)$,
which can be expressed in terms of the $\cT$-matrix by using the Dyson equation.
Let us now focus on the calculation of the $\cT$-matrix itself. For quantum impurity models, one of the most elegant ways to perform this task is to
relate it to some local correlation function that can be computed
numerically with the NRG.
For the Anderson model, the ${\cal T}$-matrix is
related to the Green's function of the $d_{\s}$ operators.~\cite{Lengreth.66, Borda.07}
For a constant and real coupling $V$, the imaginary and real parts of the 
spin-resolved $\cT$-matrix are then given by
\begin{eqnarray}
 \Im m~ \cT_{\s}(\omega) & = & -\pi\,V^2\,\cA_{\s} (\omega),\nonumber\\
 \Re e~ \cT_{\s}(\omega) & = & V^2\,{\cal P} \int \rm d\omega'\frac{\cA_{\s} (\omega')}{\omega-
 \omega'},\label{eq:ReT}
\end{eqnarray}
with $\cA_{\s}(\omega)$ the spectral function of the $d_\s$ operators
and ${\cal P}$ denoting a principal value integral.
In order to obtain reliable results for the spin-dependent spectral functions, we have employed
the density-matrix NRG.~\cite{BudapestNRG}
In addition, we have included in our calculations the full energy dependence
of $\varrho^{(0)}(E)$ for the Au(111) surface.

In Fig.~\ref{fig:spectral_function} we show the energy dependence of the 
spin-resolved spectral function $\cA_\s(\omega)$ in the vicinity of the Fermi level,
with the inset presenting its full energy dependence. Although $\varrho^{(0)}(E)$ shows a gap below $E<0.42$~meV (the bandwidth in the 
NRG calculations was fixed to $D=2.5$~eV), the two Hubbard satellites
and the Kondo peak at the Fermi level are clearly visible.
The splitting of the Kondo resonance for $B =$ 3 T is visible in $\cA_\s(\omega)$.
The applied magnetic field is not strong enough
to suppress the Kondo resonance, however it is
sufficient to produce a spin-resolved response detectable in the LMDOS.
The $\cA_\s(\omega)$ in  Fig.~\ref{fig:spectral_function}
was computed at $T=0$, but it can be argued that our findings remain valid 
as long as we are in the Kondo regime: $T < \min\{T_K, B\}$.
If the temperature increases, the Kondo peak becomes suppressed and eventually,
at high temperatures $T\gg T_K$, it is completely smeared out by thermal fluctuations.

%%%%%%%%%%%%%%%%%%%%%%%%%%%%%%%%%%%%%%%%%%%%%%%%%%%%%%%%%%%%%%%%
%%%%%%%%%%%%%%%%%%%%%%%%%%%%%%%%%%%%%%%%%%%%%%%%%%%%%%%%%%%%%%%%

\section{LMDOS: Analysis of the numerical results}\label{sec:NR}

%%%%%%%%%%%%%%%%%%%%%%%%%%%%%%%%%%%%%%%%%%%%%%%%%%%%%%%%%%%%%%%%
%%%%%%%%%%%%%%%%%%%%%%%%%%%%%%%%%%%%%%%%%%%%%%%%%%%%%%%%%%%%%%%%

In this section we shall describe how we compute the LMDOS. As discussed in Sec. \ref{sec:Intro}, the LMDOS can be related to the
single particle Green's function, see Eq. \eqref{Eq:LMDOS},
and satisfies the Dyson equation, when
expressed in terms of the $\cT-$matrix.
While this expression is somewhat cumbersome in
the chiral basis due to the presence of different form factors, in the spin space
it simplifies considerably:
\begin{multline}
  \cG_{\s\s'}(\br, \br', \omega)=\cG^{(0)}_{\s\s'}(\br, \br', \omega)+
 \sum_{\s''} \cG^{(0)}_{\s\s''}
 (\br, \br_{\rm imp}, \omega)\,\times \\
  \cT_{\s''}(\br_{\rm imp}, \omega)\,
  \cG^{(0)}_{\s''\s'}(\br_{\rm imp}, \br', \omega).
  \label{eq:Dyson}
\end{multline}
The LMDOS can be calculated from $\delta\,\cG_{\s\s'}(\br, \br, \omega) = 
\cG_{\s\s'}(\br, \br, \omega)-\cG^{(0)}_{\s\s'}(\br, \br, \omega)$, by replacing 
$\cG(\br, \br, \omega)\rightarrow \delta\cG(\br, \br, \omega)$ in Eq.~(\ref{Eq:LMDOS}).
Notice that the spin impurity acts as a simple point scatterer,  
and that in the magnetic response, only the second term in Eq.~\eqref{eq:Dyson}
gives a finite contribution. 

\begin{figure}[t]
\begin{center}
\includegraphics[width=0.49\columnwidth]{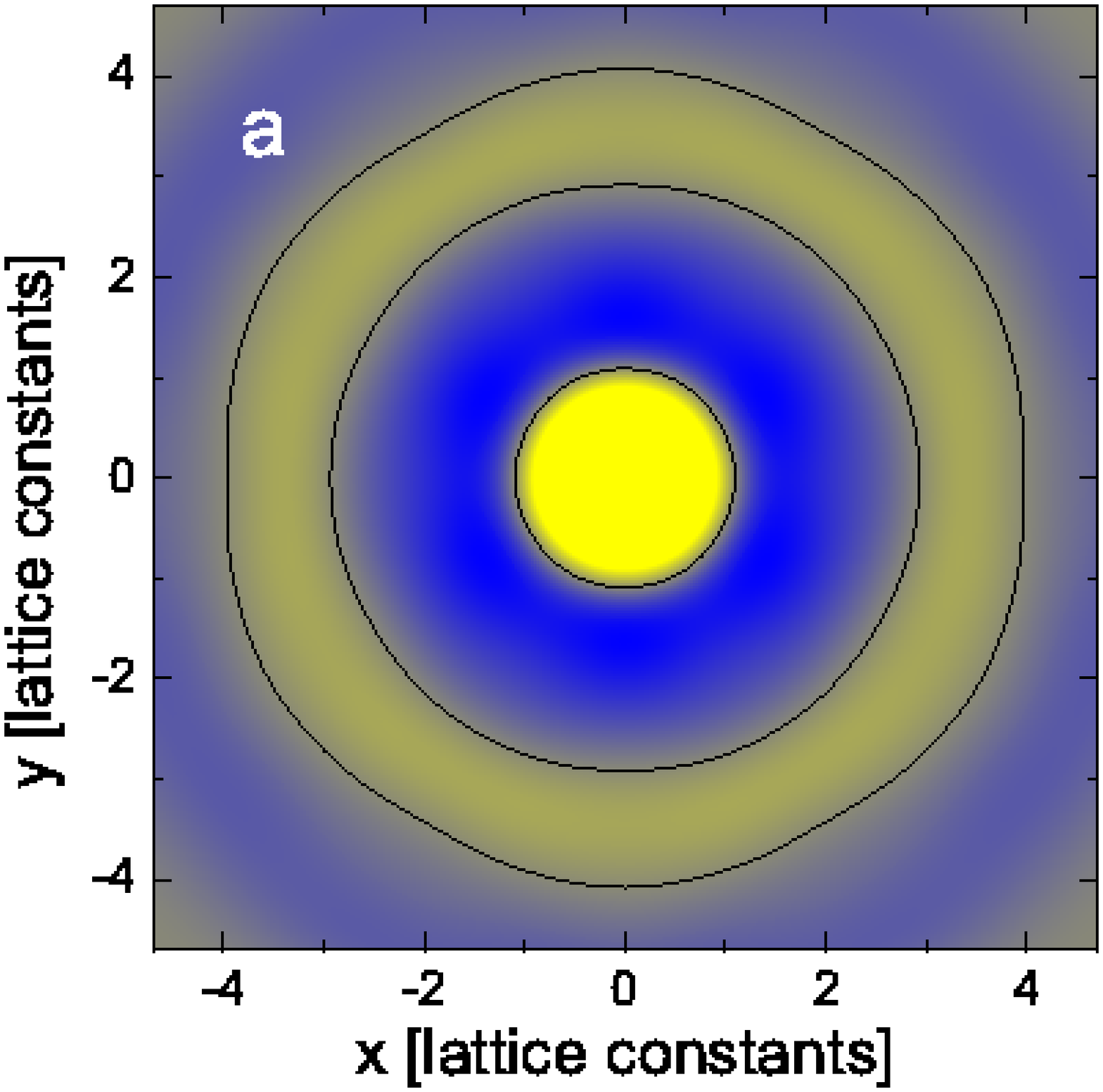}
\includegraphics[width=.49\columnwidth]{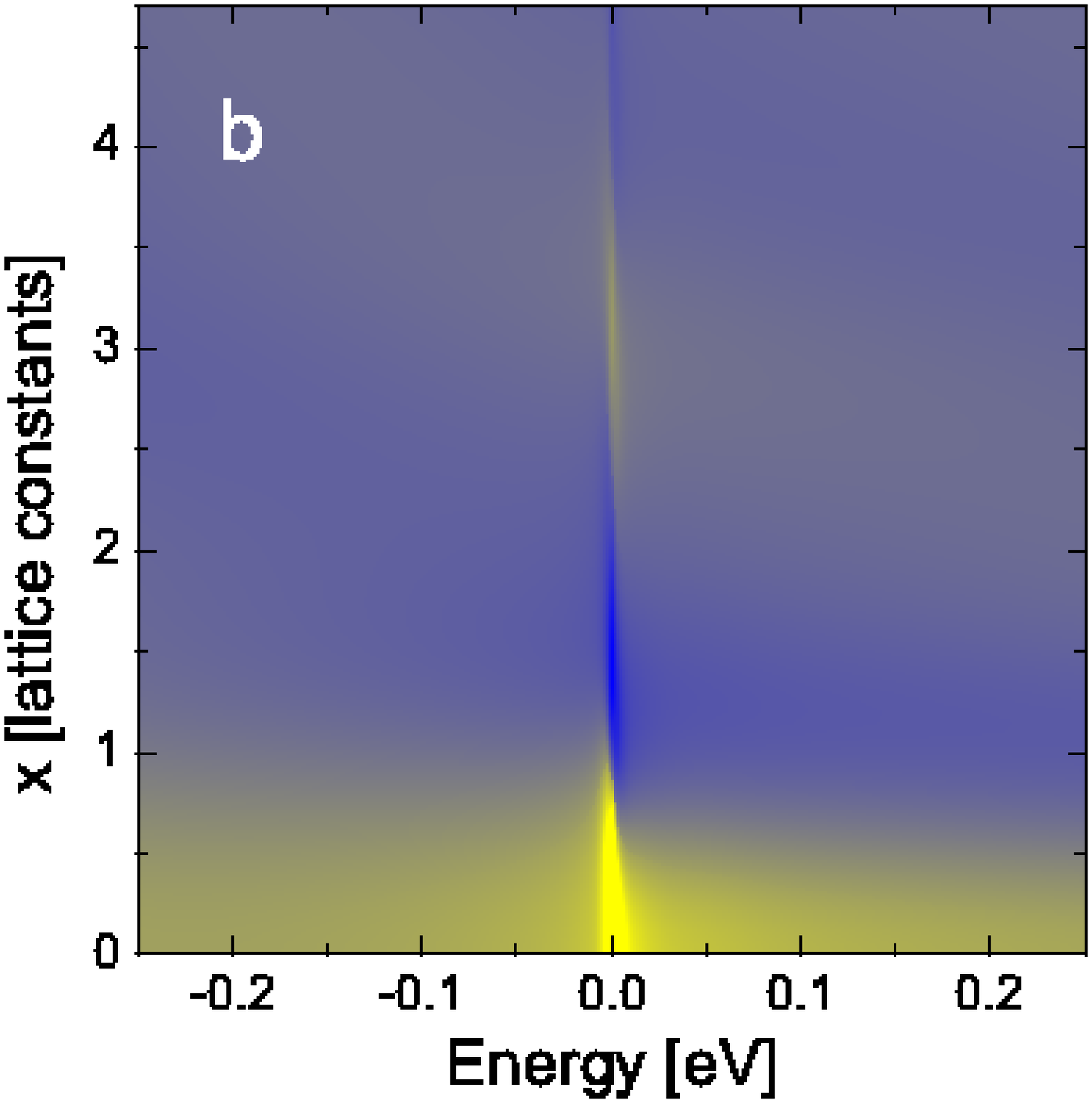}
\includegraphics[width=0.48\columnwidth]{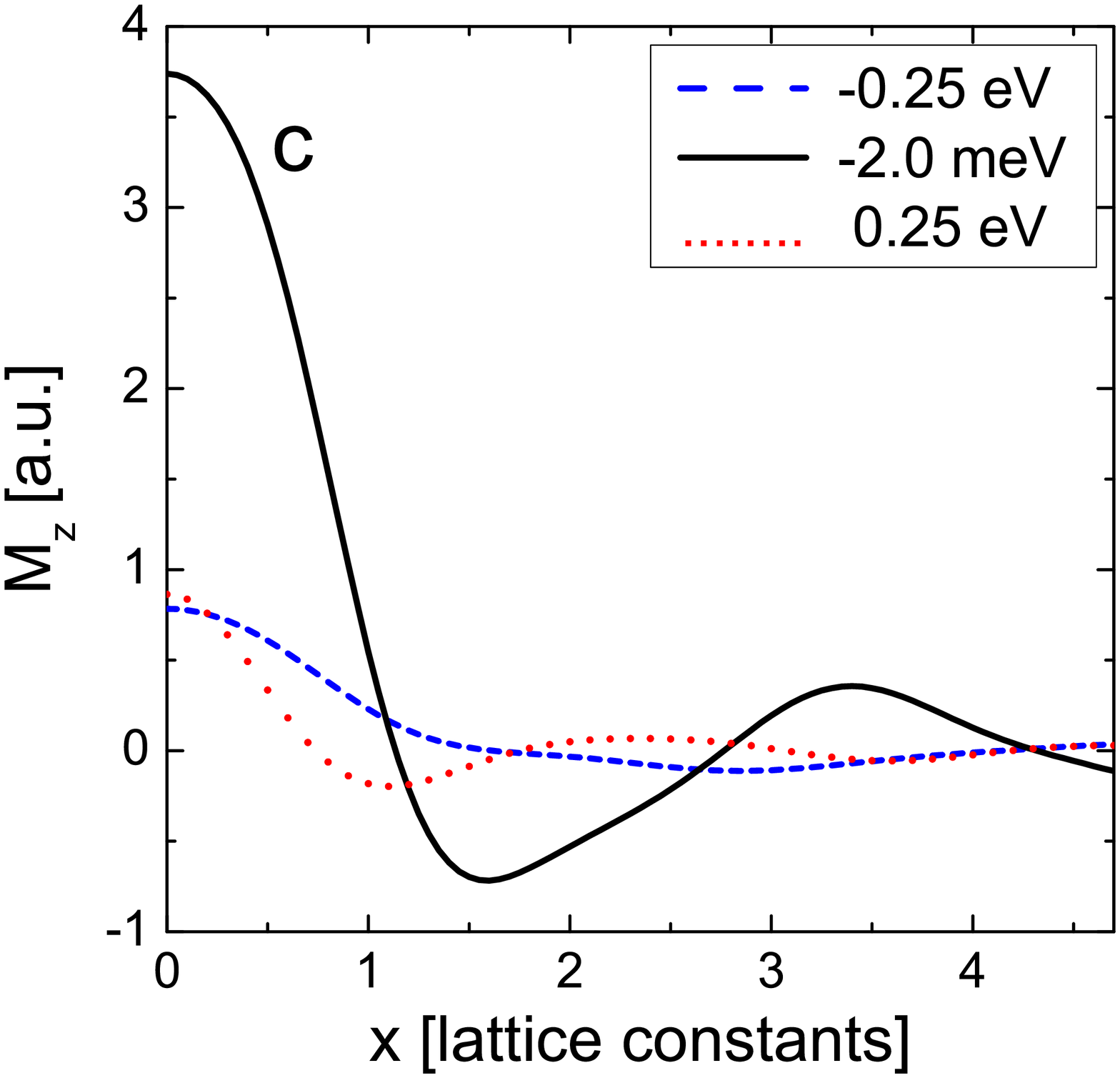}
\includegraphics[width=0.48\columnwidth]{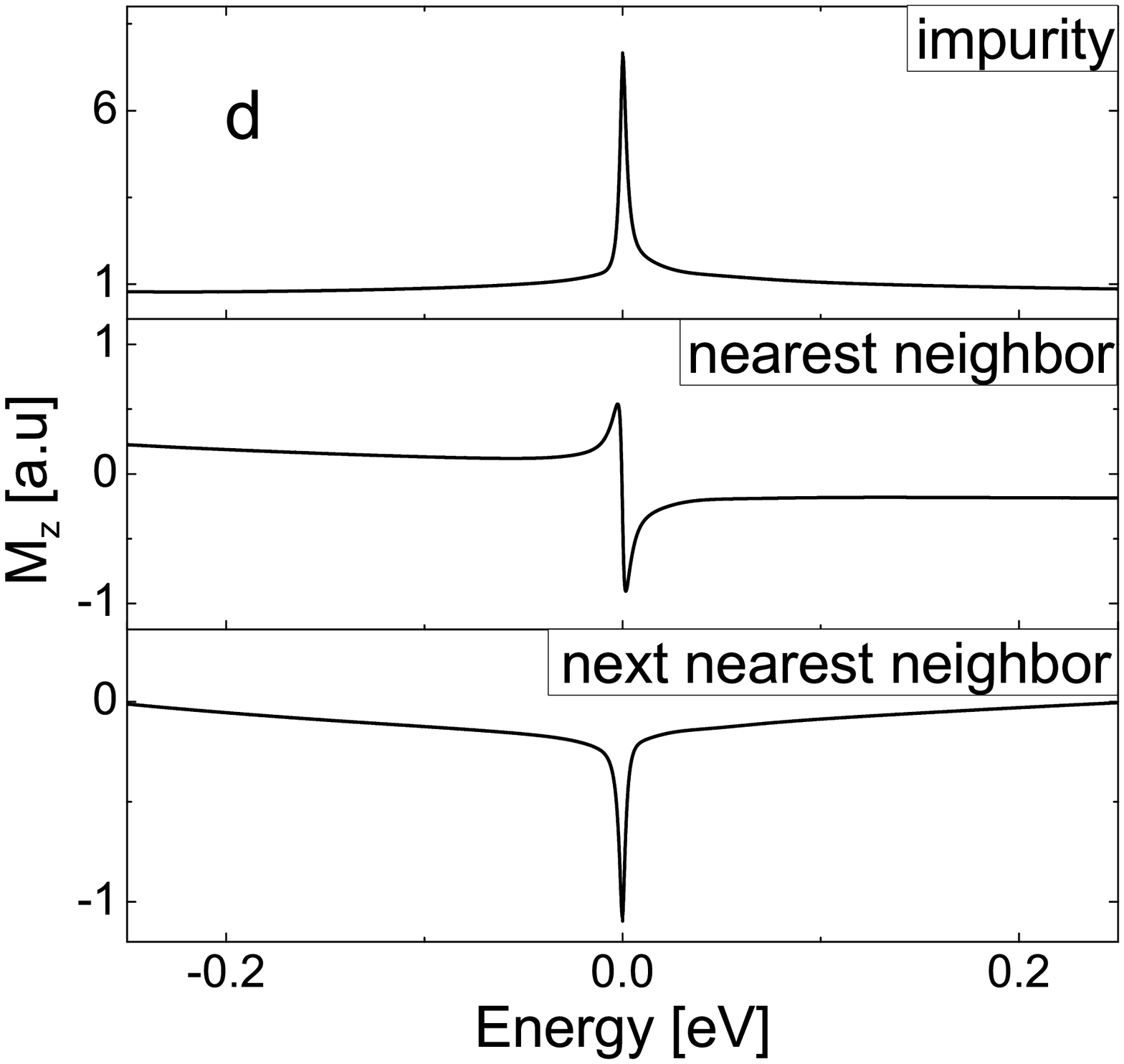}
\caption{(Color online) (a) The spatial distribution of the
  $z$th component of the LMDOS, $\rM_z$, around the Kondo impurity
  at energy $E=-2$ meV, below $E_F$ and (c) the corresponding cuts
  at different energies. The impurity is located at the center.
  The pattern formed around the magnetic impurity has a hexagonal symmetry
  and exhibits oscillations with distance from the impurity, with a period depending on the energy.
  The bright (dark) color corresponds to maximum (minimum) value of $\rM_z$.
  (b) $\rM_z$ as a function of energy and distance from the impurity along the $x$ direction, and
  (d) the energy dependence of $\rM_z(E)$ calculated at the impurity site and the two nearest-neighbor sites.}
\label{fig:Mz_plots}
\end{center}
\end{figure}

\begin{figure}[t]
\begin{center}
\includegraphics[width=0.49\columnwidth]{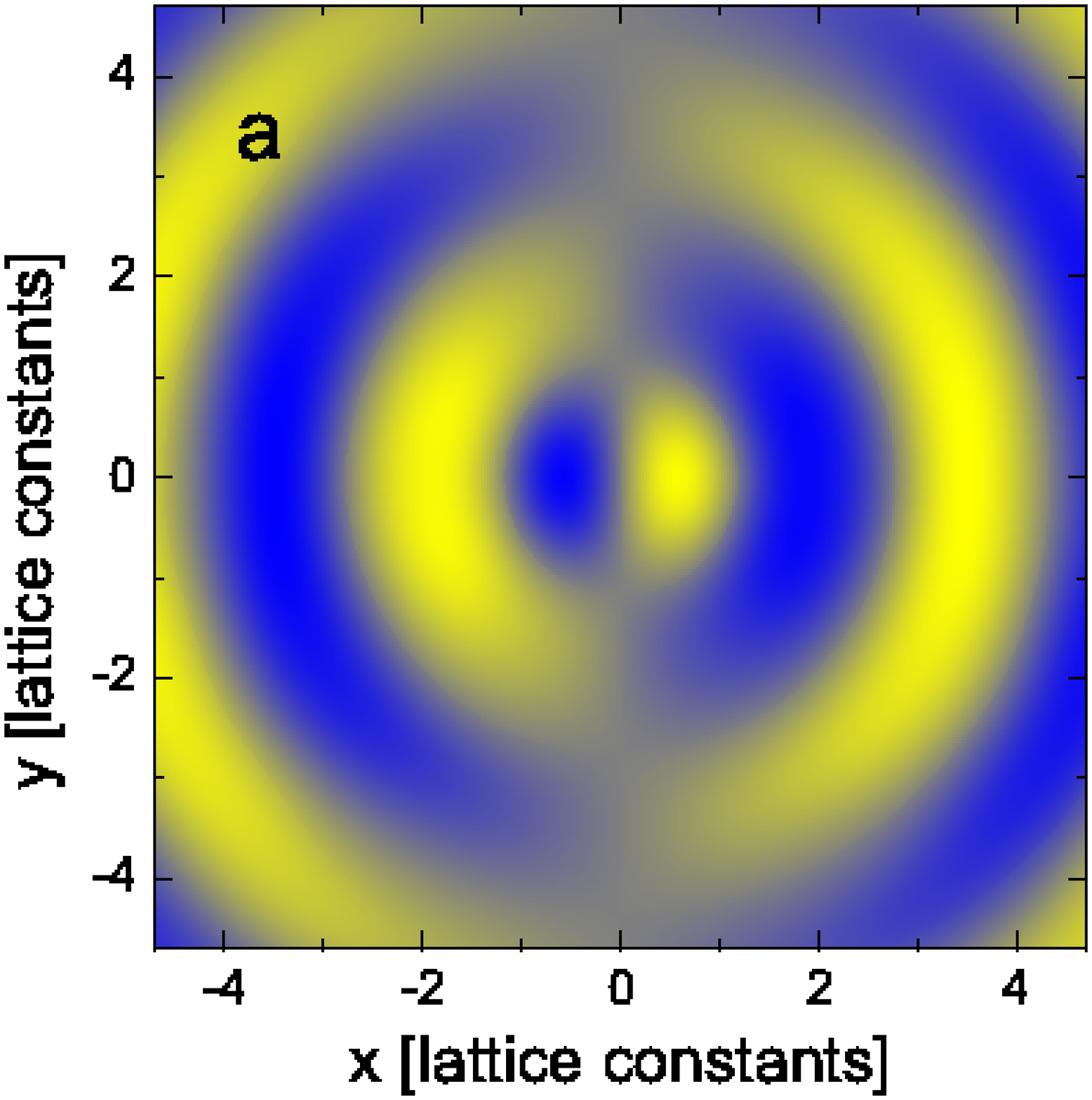}
\includegraphics[width=0.49\columnwidth]{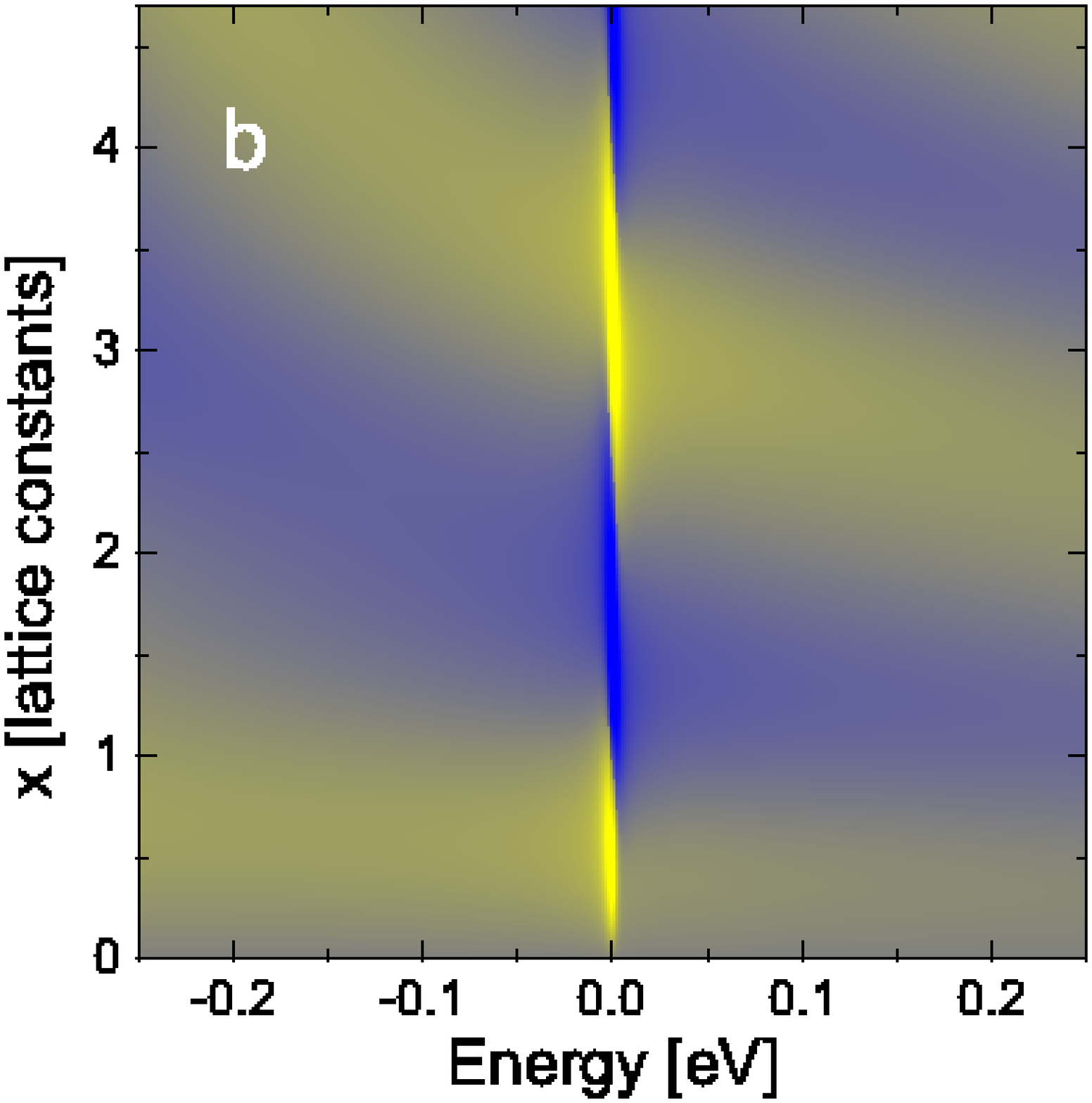}
\includegraphics[width=0.48\columnwidth]{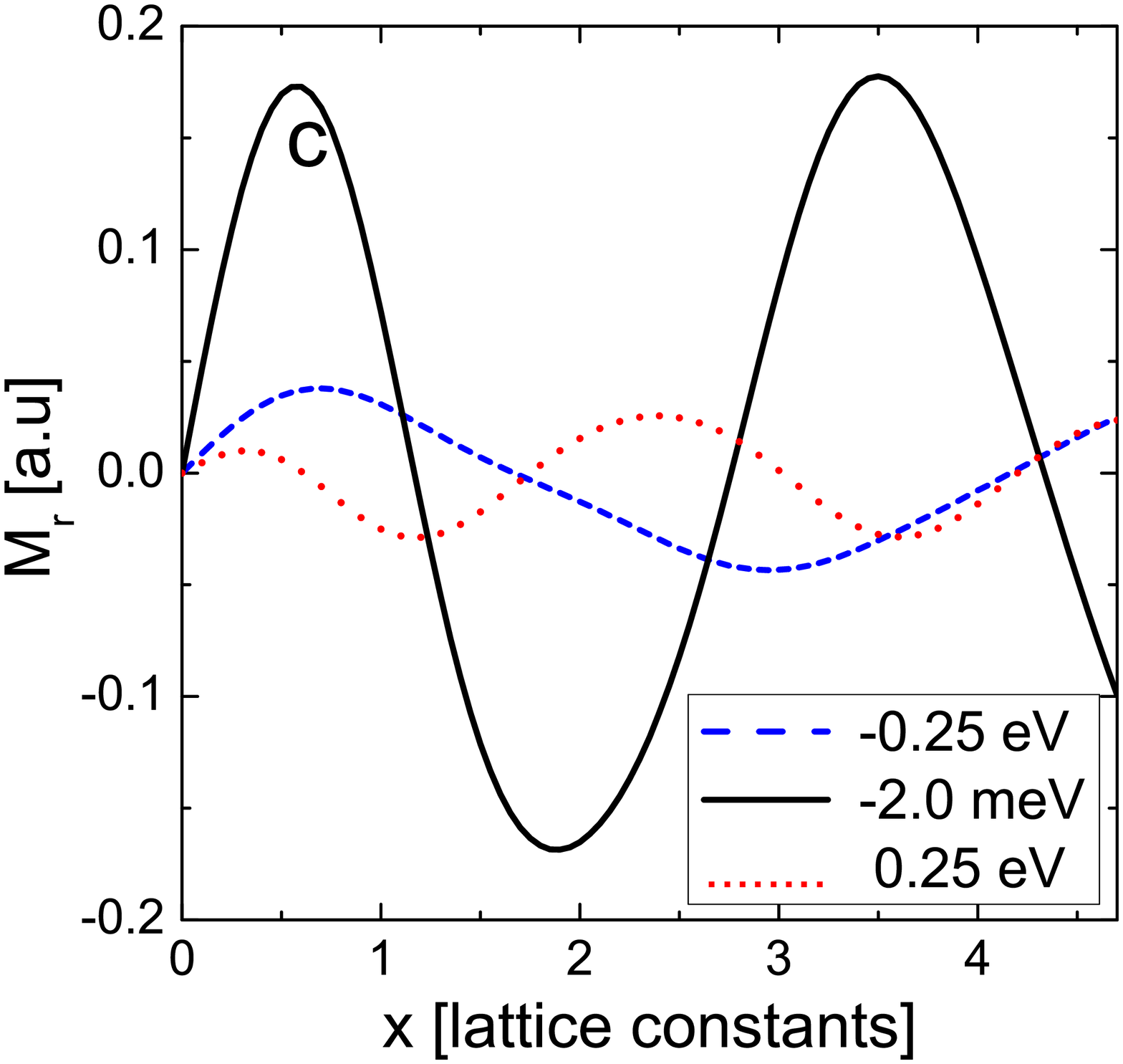}
\includegraphics[width=0.48\columnwidth]{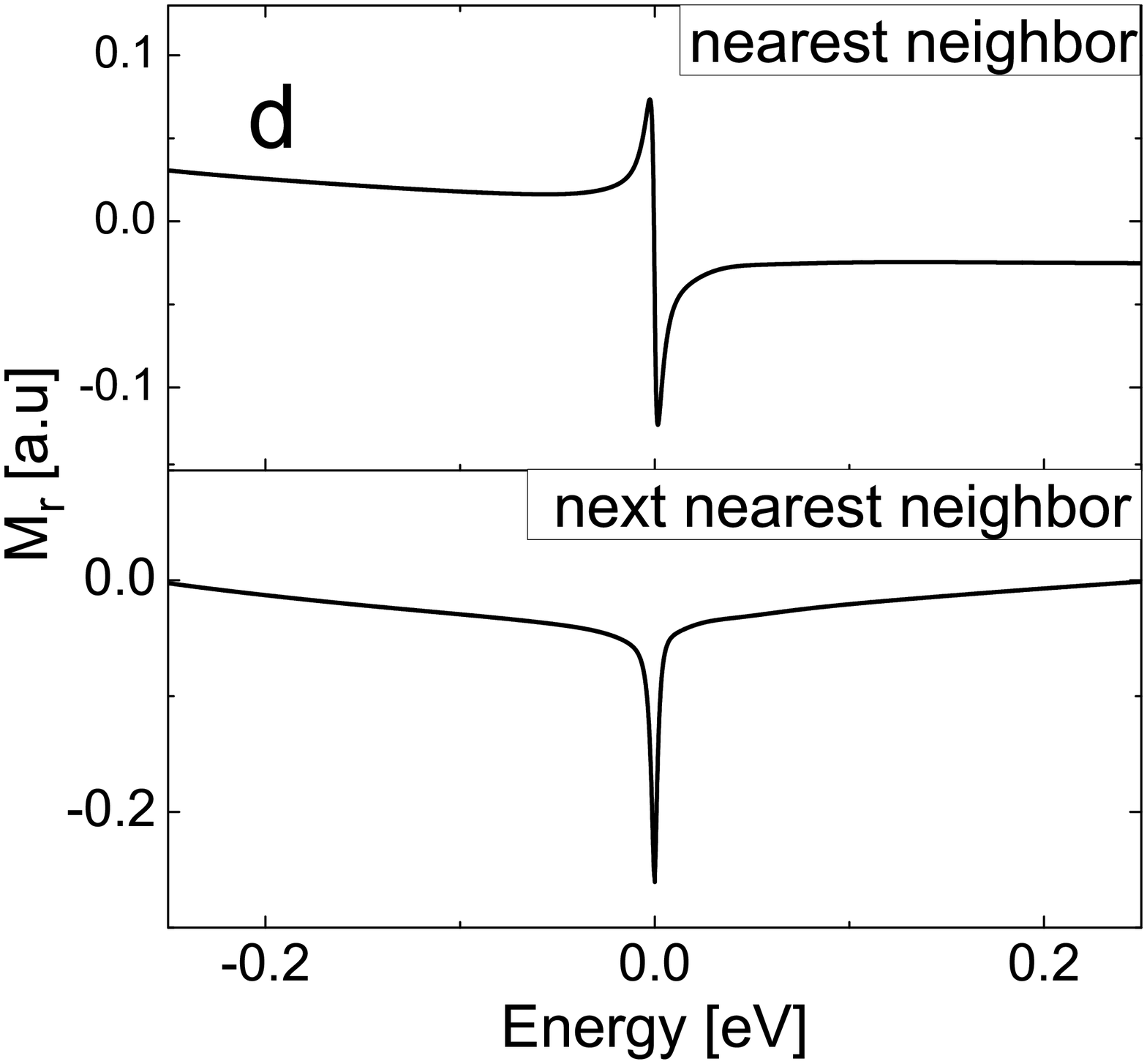}
\caption{(Color online) (a)
The spatial distribution of the in-plane x component of the LMDOS,
$\rM_x$, around the Kondo impurity at energy $E=-2$ meV, and
(c) the corresponding cuts (in the $x$ direction) at different energies.
(b) $\rM_r$ as a function
of energy and distance from the impurity along the $x$ direction,
and (d) the energy dependence of $\rM_r(\omega)$ calculated 
at the nearest and next-nearest neighbor sites.
%\rc{(Ideally Fig 4 and 5 are next to each other on the same page)}
}
\label{fig:Mr_plots}
\end{center}
\end{figure}

Within the present formalism, we are able to analyze both the spatial 
and the energy dependence of the local magnetization components. Here we will focus on $\bM_{\br}$ and
the out-of-plane, $\rM_z$ component of the LMDOS.
The spatial distribution of the $\rM_z$,
calculated at energy $E=-2$ meV, is displayed in Fig.~\ref{fig:Mz_plots}(a).
The bright (dark) areas correspond to the maxima (minima) of $\rM_z$.
One can see that its spatial dependence displays a hexagonal symmetry
with respect to the position of the magnetic impurity.
Moreover, $\rM_z$ exhibits oscillations with an energy dependent period.
This behavior is displayed in Fig.~\ref{fig:Mz_plots}(c).
When moving away from the Fermi energy, the magnitude 
of $\rM_z$ decreases and the period of the oscillations
becomes shorter as the energy is increased from negative to positive values.
This is presented in Fig.~\ref{fig:Mz_plots}(b),
which explicitly shows the energy and spatial dependence of $\rM_z$ 
when $E$ is swept across the Fermi surface. 
Close to $E_F$, the asymmetry induced by $B$
in the spin sector, together with the presence of the split Kondo resonance,
maximizes the amplitude of the local magnetization.
However, when the energy is detuned from $E_F$,
the LMDOS becomes suddenly suppressed,
leading to a Fano-like resonance, similar to the LDOS.~\cite{Miroshnichenko2010}
Moreover, the shape and magnitude of such a Fano resonance
changes as one moves away from the impurity site, see Fig.~\ref{fig:Mz_plots}(d).
One should note that within the TBA,
although we have represented the spatial distributions 
as continuous, the calculations are only valid at the atomic sites.

In a finite $B$,  the features observed in $\rM_z$
are present irrespective of the presence of the Rashba SO interaction at the surface.
On the other hand, $\rM_r$ is very sensitive to the Rashba effect.
Finite $\alpha$ implies $\rM_r\ne0$, otherwise $\rM_r$ vanishes.
The spatial distribution of $\rM_r$ around the magnetic impurity
together with its energy dependence are presented in Fig.~\ref{fig:Mr_plots}.
First of all, one can note that in the vicinity of magnetic impurity, the amplitude of $\rM_r$
is smaller by approximately one order of magnitude than the amplitude of $\rM_z$.
In fact, $\rM_r$ vanishes exactly at the impurity site, while $\rM_z$ has a maximum
there, see for example Fig.~\ref{fig:Mz_plots}(a,c) and Fig.~\ref{fig:Mr_plots}(a,c).
However, with increasing distance from the impurity, $x> 3\div 4\,a$, both 
$\rM_r$ and $\rM_z$ become comparable and, in fact,
for larger distances, the radial component can 
overtake $\rM_z$, as its decay is much slower than that of the $z$th component.
The basic properties of the spatial dependence of $\rM_r$
can be deduced from Fig.~\ref{fig:Mr_plots}(a).
One can see that $\rM_r$ is an odd function with respect to the radial distance,
and oscillates with approximately the same period as $\rM_z$.
Again, the highest amplitude of $\rM_r$ occurs for energies
around the Fermi energy due to the Kondo effect, and
as the energy increases, the period of the oscillations decreases, see Fig.~\ref{fig:Mr_plots}(b) and Fig.~\ref{fig:Mr_plots}(c).
The spatial and energy dependence of $\rM_r$ is shown in Fig.~\ref{fig:Mr_plots}(b),
while $\rM_r$ at some particular positions is displayed in panel (d). 
It can be seen that the shape of the resonance
at the nearest neighbor and the second-nearest neighbor sites
is similar to that of $\rM_z$ [cf. Fig.~\ref{fig:Mz_plots}(d)].

Finally, we would like to emphasize that $\rM_r$ 
is on one hand proportional to the strength of the SO interaction, and
on the other hand to the asymmetry of the $\cT$-matrix
components for the spin-$\uparrow$ and spin-$\downarrow$ channels.
While $\alpha$ is an intrinsic feature of the surface, the
asymmetry between the spin-$\uparrow$ and spin-$\downarrow$
channels can be changed by simply applying
an external magnetic field.
This guarantees that the topographic map of the surface
develops interference patterns in $\rM_r$ if the Rashba interaction is present.
Therefore, the measurement of the in-plane component
of the LMDOS offers an alternative way to angle resolved
photoemission spectroscopy (ARPES)~\cite{LaShell.96}
to identify surfaces with spin orbit interaction.

%%%%%%%%%%%%%%%%%%%%%%%%%%%%%%%%%%%%%%%%%%%%%%%%%%%%%%%%%%%%%%%%
%%%%%%%%%%%%%%%%%%%%%%%%%%%%%%%%%%%%%%%%%%%%%%%%%%%%%%%%%%%%%%%%

\section{Concluding Remarks}\label{sec:Conclusions}

%%%%%%%%%%%%%%%%%%%%%%%%%%%%%%%%%%%%%%%%%%%%%%%%%%%%%%%%%%%%%%%%
%%%%%%%%%%%%%%%%%%%%%%%%%%%%%%%%%%%%%%%%%%%%%%%%%%%%%%%%%%%%%%%%

In the present work we have investigated the behavior of the local magnetization
density of states around a magnetic impurity in the Kondo regime,
coupled to a metallic surface with Rashba spin orbit interaction. In order to make realistic estimates, we have 
considered a Co impurity on top of a Au(111) surface. 
This problem has been addressed using band structure calculations and the NRG
method, which allowed us to obtain reliable predictions for the LMDOS.

In particular, we have studied the spatial and energy dependence
of the LMDOS for both the radial (in-plane) and $z$th (out-of-plane)
components. We have found that the in-plane component of the LMDOS
is a pure Rashba effect.
Furthermore, it turned out that the radial component
displays oscillations with the distance from the impurity, and 
decays much slower than the $z$th component, so that
at larger distances, the in-plane component may become dominant.
Since $\rM_r$ vanishes in the absence of spin orbit interaction, 
measuring the radial component using spin-polarized STM, provides a way
to confirm/infirm the presence of the Rashba effect on surfaces.
Our observations provide thus an alternative route to investigate
the spin splitting of any surface states, first observed by
angle resolved photoemission spectroscopy~\cite{LaShell.96} in Au(111).

%%%%%%%%%%%%%%%%%%%%%%%%%%%%%%%%%%%%%%%%%%%%%%%%%%%%%%%%%%%%%%%%
%%%%%%%%%%%%%%%%%%%%%%%%%%%%%%%%%%%%%%%%%%%%%%%%%%%%%%%%%%%%%%%%
\section*{Acknowledgments}
%%%%%%%%%%%%%%%%%%%%%%%%%%%%%%%%%%%%%%%%%%%%%%%%%%%%%%%%%%%%%%%%
%%%%%%%%%%%%%%%%%%%%%%%%%%%%%%%%%%%%%%%%%%%%%%%%%%%%%%%%%%%%%%%%

This research has been supported by financial support from
UEFISCDI under French-Romanian Grant
DYMESYS (ANR 2011-IS04-001-01 and Contract
No. PN-II-ID-JRP-2011-1).
%and PN-II-ID-PCE-2012-4-0039
%and by Hungarian research funds
%OTKA and NKTH under Grant Nos.~K73361 and CNK80991,
%and the EU-NKTH GEOMDISS project.
I.W. acknowledges support from the 
`Iuventus Plus' project No. IP2011 059471 for years 2012-2014,
and the EU grant No. CIG-303 689.

%%%%%%%%%%%%%%%%%%%%%%%%%%%%%%%%%%%%%%%%%%%%%%%%%%%%%%%%%%%%%%%%
%%%%%%%%%%%%%%%%%%%%%%%%%%%%%%%%%%%%%%%%%%%%%%%%%%%%%%%%%%%%%%%%

\bibliography{references}

\begin{thebibliography}{28}%
\makeatletter
\providecommand \@ifxundefined [1]{%
 \@ifx{#1\undefined}
}%
\providecommand \@ifnum [1]{%
 \ifnum #1\expandafter \@firstoftwo
 \else \expandafter \@secondoftwo
 \fi
}%
\providecommand \@ifx [1]{%
 \ifx #1\expandafter \@firstoftwo
 \else \expandafter \@secondoftwo
 \fi
}%
\providecommand \natexlab [1]{#1}%
\providecommand \enquote  [1]{``#1''}%
\providecommand \bibnamefont  [1]{#1}%
\providecommand \bibfnamefont [1]{#1}%
\providecommand \citenamefont [1]{#1}%
\providecommand \href@noop [0]{\@secondoftwo}%
\providecommand \href [0]{\begingroup \@sanitize@url \@href}%
\providecommand \@href[1]{\@@startlink{#1}\@@href}%
\providecommand \@@href[1]{\endgroup#1\@@endlink}%
\providecommand \@sanitize@url [0]{\catcode `\\12\catcode `\$12\catcode
  `\&12\catcode `\#12\catcode `\^12\catcode `\_12\catcode `\%12\relax}%
\providecommand \@@startlink[1]{}%
\providecommand \@@endlink[0]{}%
\providecommand \url  [0]{\begingroup\@sanitize@url \@url }%
\providecommand \@url [1]{\endgroup\@href {#1}{\urlprefix }}%
\providecommand \urlprefix  [0]{URL }%
\providecommand \Eprint [0]{\href }%
\providecommand \doibase [0]{http://dx.doi.org/}%
\providecommand \selectlanguage [0]{\@gobble}%
\providecommand \bibinfo  [0]{\@secondoftwo}%
\providecommand \bibfield  [0]{\@secondoftwo}%
\providecommand \translation [1]{[#1]}%
\providecommand \BibitemOpen [0]{}%
\providecommand \bibitemStop [0]{}%
\providecommand \bibitemNoStop [0]{.\EOS\space}%
\providecommand \EOS [0]{\spacefactor3000\relax}%
\providecommand \BibitemShut  [1]{\csname bibitem#1\endcsname}%
\let\auto@bib@innerbib\@empty
%</preamble>
\bibitem [{win()}]{winkler}%
  \BibitemOpen
  \href@noop {} {\bibinfo  {journal} {R. Winkler, {\it Spin-Orbit Coupling
  effects in Two-Dimensional Electron and Hole Systems} (Springer, 2003)}\
  }\BibitemShut {NoStop}%
\bibitem [{\citenamefont {Rashba}(1960)}]{rashba}%
  \BibitemOpen
\bibfield  {journal} {  }\bibfield  {author} {\bibinfo {author} {\bibfnamefont
  {E.~I.}\ \bibnamefont {Rashba}},\ }\href@noop {} {\bibfield  {journal}
  {\bibinfo  {journal} {Sov. Phys. Solid State}\ }\textbf {\bibinfo {volume}
  {2}},\ \bibinfo {pages} {1109} (\bibinfo {year} {1960})}\BibitemShut
  {NoStop}%
\bibitem [{\citenamefont {Awschalom}\ and\ \citenamefont
  {Samarth}(2009)}]{Awschalom.09}%
  \BibitemOpen
  \bibfield  {author} {\bibinfo {author} {\bibfnamefont {D.}~\bibnamefont
  {Awschalom}}\ and\ \bibinfo {author} {\bibfnamefont {N.}~\bibnamefont
  {Samarth}},\ }\href@noop {} {\bibfield  {journal} {\bibinfo  {journal}
  {Physics}\ }\textbf {\bibinfo {volume} {2}},\ \bibinfo {pages} {50} (\bibinfo
  {year} {2009})}\BibitemShut {NoStop}%
\bibitem [{\citenamefont {LaShell}\ \emph {et~al.}(1996)\citenamefont
  {LaShell}, \citenamefont {McDougall},\ and\ \citenamefont
  {Jensen}}]{LaShell.96}%
  \BibitemOpen
  \bibfield  {author} {\bibinfo {author} {\bibfnamefont {S.}~\bibnamefont
  {LaShell}}, \bibinfo {author} {\bibfnamefont {B.~A.}\ \bibnamefont
  {McDougall}}, \ and\ \bibinfo {author} {\bibfnamefont {E.}~\bibnamefont
  {Jensen}},\ }\href@noop {} {\bibfield  {journal} {\bibinfo  {journal} {Phys.
  Rev. Lett.}\ }\textbf {\bibinfo {volume} {77}},\ \bibinfo {pages} {3419}
  (\bibinfo {year} {1996})}\BibitemShut {NoStop}%
\bibitem [{\citenamefont {Petersen}\ and\ \citenamefont
  {Hedegard}(2000)}]{Petersen.00}%
  \BibitemOpen
  \bibfield  {author} {\bibinfo {author} {\bibfnamefont {L.}~\bibnamefont
  {Petersen}}\ and\ \bibinfo {author} {\bibfnamefont {P.}~\bibnamefont
  {Hedegard}},\ }\href@noop {} {\bibfield  {journal} {\bibinfo  {journal}
  {Surface Science}\ }\textbf {\bibinfo {volume} {459}},\ \bibinfo {pages} {49
  } (\bibinfo {year} {2000})}\BibitemShut {NoStop}%
\bibitem [{\citenamefont {Lounis}\ \emph {et~al.}(2012)\citenamefont {Lounis},
  \citenamefont {Bringer},\ and\ \citenamefont {Bl\"ugel}}]{Lounis.12}%
  \BibitemOpen
  \bibfield  {author} {\bibinfo {author} {\bibfnamefont {S.}~\bibnamefont
  {Lounis}}, \bibinfo {author} {\bibfnamefont {A.}~\bibnamefont {Bringer}}, \
  and\ \bibinfo {author} {\bibfnamefont {S.}~\bibnamefont {Bl\"ugel}},\
  }\href@noop {} {\bibfield  {journal} {\bibinfo  {journal} {Phys. Rev. Lett.}\
  }\textbf {\bibinfo {volume} {108}},\ \bibinfo {pages} {207202} (\bibinfo
  {year} {2012})}\BibitemShut {NoStop}%
\bibitem [{\citenamefont {Kondo}(1964)}]{Kondo.64}%
  \BibitemOpen
  \bibfield  {author} {\bibinfo {author} {\bibfnamefont {J.}~\bibnamefont
  {Kondo}},\ }\href@noop {} {\bibfield  {journal} {\bibinfo  {journal}
  {Progress of Theoretical Physics}\ }\textbf {\bibinfo {volume} {32}},\
  \bibinfo {pages} {37} (\bibinfo {year} {1964})}\BibitemShut {NoStop}%
\bibitem [{hew()}]{hewson}%
  \BibitemOpen
  \href@noop {} {\bibinfo  {journal} {A. C. Hewson, {\it The Kondo Problem to
  Heavy Fermions} (Cambridge University Press, Cambridge, 1993)}\ }\BibitemShut
  {NoStop}%
\bibitem [{\citenamefont {Madhavan}\ \emph {et~al.}(1998)\citenamefont
  {Madhavan}, \citenamefont {Chen}, \citenamefont {Jamneala}, \citenamefont
  {Crommie},\ and\ \citenamefont {Wingreen}}]{Madhavan.98}%
  \BibitemOpen
\bibfield  {journal} {  }\bibfield  {author} {\bibinfo {author} {\bibfnamefont
  {V.}~\bibnamefont {Madhavan}}, \bibinfo {author} {\bibfnamefont
  {W.}~\bibnamefont {Chen}}, \bibinfo {author} {\bibfnamefont {T.}~\bibnamefont
  {Jamneala}}, \bibinfo {author} {\bibfnamefont {M.~F.}\ \bibnamefont
  {Crommie}}, \ and\ \bibinfo {author} {\bibfnamefont {N.~S.}\ \bibnamefont
  {Wingreen}},\ }\href@noop {} {\bibfield  {journal} {\bibinfo  {journal}
  {Science}\ }\textbf {\bibinfo {volume} {280}},\ \bibinfo {pages} {567}
  (\bibinfo {year} {1998})}\BibitemShut {NoStop}%
\bibitem [{\citenamefont {Madhavan}\ \emph {et~al.}(2001)\citenamefont
  {Madhavan}, \citenamefont {Chen}, \citenamefont {Jamneala}, \citenamefont
  {Crommie},\ and\ \citenamefont {Wingreen}}]{Madhavan.01}%
  \BibitemOpen
  \bibfield  {author} {\bibinfo {author} {\bibfnamefont {V.}~\bibnamefont
  {Madhavan}}, \bibinfo {author} {\bibfnamefont {W.}~\bibnamefont {Chen}},
  \bibinfo {author} {\bibfnamefont {T.}~\bibnamefont {Jamneala}}, \bibinfo
  {author} {\bibfnamefont {M.~F.}\ \bibnamefont {Crommie}}, \ and\ \bibinfo
  {author} {\bibfnamefont {N.~S.}\ \bibnamefont {Wingreen}},\ }\href@noop {}
  {\bibfield  {journal} {\bibinfo  {journal} {Phys. Rev. B}\ }\textbf {\bibinfo
  {volume} {64}},\ \bibinfo {pages} {165412} (\bibinfo {year}
  {2001})}\BibitemShut {NoStop}%
\bibitem [{\citenamefont {Liu}\ \emph {et~al.}(2008)\citenamefont {Liu},
  \citenamefont {Chen},\ and\ \citenamefont {Chang}}]{Liu.08}%
  \BibitemOpen
  \bibfield  {author} {\bibinfo {author} {\bibfnamefont {M.-H.}\ \bibnamefont
  {Liu}}, \bibinfo {author} {\bibfnamefont {S.-H.}\ \bibnamefont {Chen}}, \
  and\ \bibinfo {author} {\bibfnamefont {C.-R.}\ \bibnamefont {Chang}},\
  }\href@noop {} {\bibfield  {journal} {\bibinfo  {journal} {Phys. Rev. B}\
  }\textbf {\bibinfo {volume} {78}},\ \bibinfo {pages} {195413} (\bibinfo
  {year} {2008})}\BibitemShut {NoStop}%
\bibitem [{\citenamefont {Anderson}(1961)}]{Anderson.61}%
  \BibitemOpen
  \bibfield  {author} {\bibinfo {author} {\bibfnamefont {P.~W.}\ \bibnamefont
  {Anderson}},\ }\href@noop {} {\bibfield  {journal} {\bibinfo  {journal}
  {Phys. Rev.}\ }\textbf {\bibinfo {volume} {124}},\ \bibinfo {pages} {41}
  (\bibinfo {year} {1961})}\BibitemShut {NoStop}%
\bibitem [{\citenamefont {Wilson}(1975)}]{Wilson.75}%
  \BibitemOpen
  \bibfield  {author} {\bibinfo {author} {\bibfnamefont {K.~G.}\ \bibnamefont
  {Wilson}},\ }\href@noop {} {\bibfield  {journal} {\bibinfo  {journal} {Rev.
  Mod. Phys.}\ }\textbf {\bibinfo {volume} {47}},\ \bibinfo {pages} {773}
  (\bibinfo {year} {1975})}\BibitemShut {NoStop}%
\bibitem [{\citenamefont {Knorr}\ \emph {et~al.}(2002)\citenamefont {Knorr},
  \citenamefont {Schneider}, \citenamefont {Diekh\"oner}, \citenamefont
  {Wahl},\ and\ \citenamefont {Kern}}]{Knorr.02}%
  \BibitemOpen
  \bibfield  {author} {\bibinfo {author} {\bibfnamefont {N.}~\bibnamefont
  {Knorr}}, \bibinfo {author} {\bibfnamefont {M.~A.}\ \bibnamefont
  {Schneider}}, \bibinfo {author} {\bibfnamefont {L.}~\bibnamefont
  {Diekh\"oner}}, \bibinfo {author} {\bibfnamefont {P.}~\bibnamefont {Wahl}}, \
  and\ \bibinfo {author} {\bibfnamefont {K.}~\bibnamefont {Kern}},\ }\href@noop
  {} {\bibfield  {journal} {\bibinfo  {journal} {Phys. Rev. Lett.}\ }\textbf
  {\bibinfo {volume} {88}},\ \bibinfo {pages} {096804} (\bibinfo {year}
  {2002})}\BibitemShut {NoStop}%
\bibitem [{\citenamefont {Schneider}\ \emph {et~al.}(2005)\citenamefont
  {Schneider}, \citenamefont {Vitali}, \citenamefont {Wahl}, \citenamefont
  {Knorr}, \citenamefont {Diekhauner}, \citenamefont {Wittich}, \citenamefont
  {Vogelgesang},\ and\ \citenamefont {Kern}}]{Schneider.05}%
  \BibitemOpen
  \bibfield  {author} {\bibinfo {author} {\bibfnamefont {M.}~\bibnamefont
  {Schneider}}, \bibinfo {author} {\bibfnamefont {L.}~\bibnamefont {Vitali}},
  \bibinfo {author} {\bibfnamefont {P.}~\bibnamefont {Wahl}}, \bibinfo {author}
  {\bibfnamefont {N.}~\bibnamefont {Knorr}}, \bibinfo {author} {\bibfnamefont
  {L.}~\bibnamefont {Diekhauner}}, \bibinfo {author} {\bibfnamefont
  {G.}~\bibnamefont {Wittich}}, \bibinfo {author} {\bibfnamefont
  {M.}~\bibnamefont {Vogelgesang}}, \ and\ \bibinfo {author} {\bibfnamefont
  {K.}~\bibnamefont {Kern}},\ }\href@noop {} {\bibfield  {journal} {\bibinfo
  {journal} {Applied Physics A: Materials Science and Processing}\ }\textbf
  {\bibinfo {volume} {80}},\ \bibinfo {pages} {937} (\bibinfo {year}
  {2005})}\BibitemShut {NoStop}%
\bibitem [{\citenamefont {Wiesendanger}(2009)}]{Wiesendanger.09}%
  \BibitemOpen
  \bibfield  {author} {\bibinfo {author} {\bibfnamefont {R.}~\bibnamefont
  {Wiesendanger}},\ }\href@noop {} {\bibfield  {journal} {\bibinfo  {journal}
  {Rev. Mod. Phys.}\ }\textbf {\bibinfo {volume} {81}},\ \bibinfo {pages}
  {1495} (\bibinfo {year} {2009})}\BibitemShut {NoStop}%
\bibitem [{\citenamefont {Chen}\ \emph {et~al.}(1998)\citenamefont {Chen},
  \citenamefont {Madhavan}, \citenamefont {Jamneala},\ and\ \citenamefont
  {Crommie}}]{Chen.98}%
  \BibitemOpen
  \bibfield  {author} {\bibinfo {author} {\bibfnamefont {W.}~\bibnamefont
  {Chen}}, \bibinfo {author} {\bibfnamefont {V.}~\bibnamefont {Madhavan}},
  \bibinfo {author} {\bibfnamefont {T.}~\bibnamefont {Jamneala}}, \ and\
  \bibinfo {author} {\bibfnamefont {M.~F.}\ \bibnamefont {Crommie}},\
  }\href@noop {} {\bibfield  {journal} {\bibinfo  {journal} {Phys. Rev. Lett.}\
  }\textbf {\bibinfo {volume} {80}},\ \bibinfo {pages} {1469} (\bibinfo {year}
  {1998})}\BibitemShut {NoStop}%
\bibitem [{\citenamefont {Takeuchi}\ \emph {et~al.}(1991)\citenamefont
  {Takeuchi}, \citenamefont {Chan},\ and\ \citenamefont {Ho}}]{Takeuri.91}%
  \BibitemOpen
  \bibfield  {author} {\bibinfo {author} {\bibfnamefont {N.}~\bibnamefont
  {Takeuchi}}, \bibinfo {author} {\bibfnamefont {C.~T.}\ \bibnamefont {Chan}},
  \ and\ \bibinfo {author} {\bibfnamefont {K.~M.}\ \bibnamefont {Ho}},\
  }\href@noop {} {\bibfield  {journal} {\bibinfo  {journal} {Phys. Rev. B}\
  }\textbf {\bibinfo {volume} {43}},\ \bibinfo {pages} {13899} (\bibinfo {year}
  {1991})}\BibitemShut {NoStop}%
\bibitem [{\citenamefont {\'Ujs\'aghy}\ \emph {et~al.}(2000)\citenamefont
  {\'Ujs\'aghy}, \citenamefont {Kroha}, \citenamefont {Szunyogh},\ and\
  \citenamefont {Zawadowski}}]{Ujsaghy.00}%
  \BibitemOpen
  \bibfield  {author} {\bibinfo {author} {\bibfnamefont {O.}~\bibnamefont
  {\'Ujs\'aghy}}, \bibinfo {author} {\bibfnamefont {J.}~\bibnamefont {Kroha}},
  \bibinfo {author} {\bibfnamefont {L.}~\bibnamefont {Szunyogh}}, \ and\
  \bibinfo {author} {\bibfnamefont {A.}~\bibnamefont {Zawadowski}},\
  }\href@noop {} {\bibfield  {journal} {\bibinfo  {journal} {Phys. Rev. Lett.}\
  }\textbf {\bibinfo {volume} {85}},\ \bibinfo {pages} {2557} (\bibinfo {year}
  {2000})}\BibitemShut {NoStop}%
\bibitem [{\citenamefont {Costi}(2000)}]{Costi.2000}%
  \BibitemOpen
  \bibfield  {author} {\bibinfo {author} {\bibfnamefont {T.~A.}\ \bibnamefont
  {Costi}},\ }\href@noop {} {\bibfield  {journal} {\bibinfo  {journal} {Phys.
  Rev. Lett.}\ }\textbf {\bibinfo {volume} {85}},\ \bibinfo {pages} {1504}
  (\bibinfo {year} {2000})}\BibitemShut {NoStop}%
\bibitem [{\citenamefont {Kretinin}\ \emph {et~al.}(2011)\citenamefont
  {Kretinin}, \citenamefont {Shtrikman}, \citenamefont {Goldhaber-Gordon},
  \citenamefont {Hanl}, \citenamefont {Weichselbaum}, \citenamefont {von
  Delft}, \citenamefont {Costi},\ and\ \citenamefont {Mahalu}}]{Kretinin.2011}%
  \BibitemOpen
  \bibfield  {author} {\bibinfo {author} {\bibfnamefont {A.~V.}\ \bibnamefont
  {Kretinin}}, \bibinfo {author} {\bibfnamefont {H.}~\bibnamefont {Shtrikman}},
  \bibinfo {author} {\bibfnamefont {D.}~\bibnamefont {Goldhaber-Gordon}},
  \bibinfo {author} {\bibfnamefont {M.}~\bibnamefont {Hanl}}, \bibinfo {author}
  {\bibfnamefont {A.}~\bibnamefont {Weichselbaum}}, \bibinfo {author}
  {\bibfnamefont {J.}~\bibnamefont {von Delft}}, \bibinfo {author}
  {\bibfnamefont {T.}~\bibnamefont {Costi}}, \ and\ \bibinfo {author}
  {\bibfnamefont {D.}~\bibnamefont {Mahalu}},\ }\href@noop {} {\bibfield
  {journal} {\bibinfo  {journal} {Phys. Rev. B}\ }\textbf {\bibinfo {volume}
  {84}},\ \bibinfo {pages} {245316} (\bibinfo {year} {2011})}\BibitemShut
  {NoStop}%
\bibitem [{\citenamefont {Wei}\ \emph {et~al.}(1989)\citenamefont {Wei},
  \citenamefont {Rosenbaum},\ and\ \citenamefont {Bergmann}}]{Wei.89}%
  \BibitemOpen
  \bibfield  {author} {\bibinfo {author} {\bibfnamefont {W.}~\bibnamefont
  {Wei}}, \bibinfo {author} {\bibfnamefont {R.}~\bibnamefont {Rosenbaum}}, \
  and\ \bibinfo {author} {\bibfnamefont {G.}~\bibnamefont {Bergmann}},\
  }\href@noop {} {\bibfield  {journal} {\bibinfo  {journal} {Phys. Rev. B}\
  }\textbf {\bibinfo {volume} {39}},\ \bibinfo {pages} {4568} (\bibinfo {year}
  {1989})}\BibitemShut {NoStop}%
\bibitem [{\citenamefont {Bulla}\ \emph {et~al.}(2008)\citenamefont {Bulla},
  \citenamefont {Costi},\ and\ \citenamefont {Pruschke}}]{Bulla.2008}%
  \BibitemOpen
  \bibfield  {author} {\bibinfo {author} {\bibfnamefont {R.}~\bibnamefont
  {Bulla}}, \bibinfo {author} {\bibfnamefont {T.~A.}\ \bibnamefont {Costi}}, \
  and\ \bibinfo {author} {\bibfnamefont {T.}~\bibnamefont {Pruschke}},\
  }\href@noop {} {\bibfield  {journal} {\bibinfo  {journal} {Rev. Mod. Phys.}\
  }\textbf {\bibinfo {volume} {80}},\ \bibinfo {pages} {395} (\bibinfo {year}
  {2008})}\BibitemShut {NoStop}%
\bibitem [{\citenamefont {Krishna-murthy}\ \emph {et~al.}(1980)\citenamefont
  {Krishna-murthy}, \citenamefont {Wilkins},\ and\ \citenamefont
  {Wilson}}]{Krishnamurthy.80}%
  \BibitemOpen
  \bibfield  {author} {\bibinfo {author} {\bibfnamefont {H.~R.}\ \bibnamefont
  {Krishna-murthy}}, \bibinfo {author} {\bibfnamefont {J.~W.}\ \bibnamefont
  {Wilkins}}, \ and\ \bibinfo {author} {\bibfnamefont {K.~G.}\ \bibnamefont
  {Wilson}},\ }\href@noop {} {\bibfield  {journal} {\bibinfo  {journal} {Phys.
  Rev. B}\ }\textbf {\bibinfo {volume} {21}},\ \bibinfo {pages} {1003}
  (\bibinfo {year} {1980})}\BibitemShut {NoStop}%
\bibitem [{\citenamefont {Langreth}(1966)}]{Lengreth.66}%
  \BibitemOpen
  \bibfield  {author} {\bibinfo {author} {\bibfnamefont {D.~C.}\ \bibnamefont
  {Langreth}},\ }\href@noop {} {\bibfield  {journal} {\bibinfo  {journal}
  {Phys. Rev.}\ }\textbf {\bibinfo {volume} {150}},\ \bibinfo {pages} {516}
  (\bibinfo {year} {1966})}\BibitemShut {NoStop}%
\bibitem [{\citenamefont {Borda}\ \emph {et~al.}(2007)\citenamefont {Borda},
  \citenamefont {Fritz}, \citenamefont {Andrei},\ and\ \citenamefont
  {Zar\'and}}]{Borda.07}%
  \BibitemOpen
  \bibfield  {author} {\bibinfo {author} {\bibfnamefont {L.}~\bibnamefont
  {Borda}}, \bibinfo {author} {\bibfnamefont {L.}~\bibnamefont {Fritz}},
  \bibinfo {author} {\bibfnamefont {N.}~\bibnamefont {Andrei}}, \ and\ \bibinfo
  {author} {\bibfnamefont {G.}~\bibnamefont {Zar\'and}},\ }\href@noop {}
  {\bibfield  {journal} {\bibinfo  {journal} {Phys. Rev. B}\ }\textbf {\bibinfo
  {volume} {75}},\ \bibinfo {pages} {235112} (\bibinfo {year}
  {2007})}\BibitemShut {NoStop}%
\bibitem [{Bud()}]{BudapestNRG}%
  \BibitemOpen
  \href@noop {} {\bibinfo  {journal} {We used the open-access Budapest Flexible
  DM-NRG code, http://www.phy.bme.hu/\~{}dmnrg/; O. Legeza, C. P. Moca, A. I.
  T\'{o}th, I. Weymann, G. Zar\'{a}nd, arXiv:0809.3143 (2008) (unpublished)}\
  }\BibitemShut {NoStop}%
\bibitem [{\citenamefont {Miroshnichenko}\ \emph {et~al.}(2010)\citenamefont
  {Miroshnichenko}, \citenamefont {Flach},\ and\ \citenamefont
  {Kivshar}}]{Miroshnichenko2010}%
  \BibitemOpen
\bibfield  {journal} {  }\bibfield  {author} {\bibinfo {author} {\bibfnamefont
  {A.~E.}\ \bibnamefont {Miroshnichenko}}, \bibinfo {author} {\bibfnamefont
  {S.}~\bibnamefont {Flach}}, \ and\ \bibinfo {author} {\bibfnamefont {Y.~S.}\
  \bibnamefont {Kivshar}},\ }\href@noop {} {\bibfield  {journal} {\bibinfo
  {journal} {Rev. Mod. Phys.}\ }\textbf {\bibinfo {volume} {82}},\ \bibinfo
  {pages} {2257} (\bibinfo {year} {2010})}\BibitemShut {NoStop}%
\end{thebibliography}%

\end{document}